\documentclass{jfm}
\usepackage{graphicx}
\usepackage{epstopdf,epsfig,amsmath}
\usepackage{mathtools}
\usepackage{mathrsfs}
\usepackage{xcolor}
\usepackage{lineno}
\usepackage{hyperref}
\newcommand{\St}{S\kern-0.04em t\,}
\newcommand{\Sts}{S\kern-0.04em t}
\newcommand{\Ca}{C\kern-0.04em a\,}
\newcommand{\Cas}{C\kern-0.04em a}

\shorttitle{Decreasing contact angles at accelerating contact lines}
\shortauthor{Carlos A. Galeano-Rios}

\title{Decreasing contact angles at accelerating three-phase moving contact lines.}

\author{Carlos A. Galeano-Rios\aff{1}
  \corresp{\email{galeanca@bham.ac.uk}}}

\affiliation{\aff{1}School of Mathematics, University of Birmingham,
Birmingham, B15 2TT, United Kingdom}

\begin{document}

\maketitle

\begin{abstract}
We consider a liquid drop placed on a smooth homogeneous solid substrate as it spreads from rest to its eventual equilibrium state. The problem is studied numerically in the framework of a model where the contact angle formed by the drop's free surface with the substrate is not prescribed as a function of the contact-line speed and has to be found as part of the solution. It is shown that in this spontaneous spreading, as the drop starts moving and the contact line accelerates, the dynamic contact angle decreases with the contact-line speed, which is in line with what experimental observations on spontaneous wetting report, though it is contrary to what is assumed in most models aimed at describing experiments on the ``forced'' spreading. Nontrivial aspects of the implementation of the model in a finite-element-based algorithm are discussed in detail.
\end{abstract}


\section{Introduction}
% Processes in which a liquid spreads over a solid are common in many industrial applications (e.g. coating of surfaces and additive manufacturing); consequently, interest in modelling dynamic wetting has been fuelled by its industrial relevance \citep{SnoeijerAndAndreotti2013,RalstonEtAl2008}. Theoretical motivations have also been of importance in driving the field. Following the work of \citet{HuhAndScriven1971}, who showed that the no-slip condition leads to a non-integrable singularity in the stress at the contact line, it was clear that a new formulation of the problem; in particular, one that relaxed the no-slip condition near the contact line, was necessary. Moreover, the contact angle still needed to be determined or input into the model.

The relationship between the dynamic contact angle and the contact-line speed in dynamic wetting has been intensively studied both experimentally (e.g. \citep{Hoffman1974,NganAndDussan1982,SeeberghAndBerg1992,SikaloEtAl2005}) and theoretically (e.g. \citet{Ranabothu2EtAl2005,RenAndE2007,LegendreAndMaglio2013}). In theoretical studies, almost invariably the contact angle is assumed to be a function of the contact-line speed, which assumes the contact angle increases as the speed increases. This function, assumed the same for all flows, is used as a fixed input into the models.

However, recent experiments \citep{KarimEtAl2016} have shown that the  contact  angle-vs-speed  relationship  is quantitatively  different  for  the  so-called  ``spontaneous''  and ``forced''  contact-line  motion,  and  the  discrepancy  goes  beyond  the  experimental  error. Furthermore, experiments by \citet{BayerAndMegaridis2006} also show that the dynamic contact angle can decrease as the contact-line speed increases with the authors coming to the same conclusion that no one-to-one map from the contact-line velocity to  the dynamic contact angle can account for all types of flows and even for different regimes of qualitatively the same flow. The situation has been qualitatively summarised in a recent review \citep{Shikhmurzaev2020}.

The direct measurements of the dependence of the dynamic contact angle on contact line velocity \citep{KarimEtAl2016,BayerAndMegaridis2006}, as well as earlier experiments on the flow dependence of the contact angle \citep{RuijterEtAl1999} and their analysis, \citep{DaviesEtAL2006} indicate that the dynamic contact angle should be an outcome, part of the solution, rather than an input in the form of a given function into the model. Such sensitivity of the contact angle and hence the entire free surface shape to the flow make any realistic mathematical model much more complex than the simpler ‘slip models’ explored so far (see \citet{Shikhmurzaev2020} for a review) and its adequate numerical implementation a formidable problem in its own right.

The only known model to date that treats the contact angle as an outcome rather than an input is the model introduced in \citet{Shikhmurzaev1993} and further developed in \citet{Shikhmurzaev2007,SprittlesAndShikhmurzaev2012}, where dynamic wetting is described as a process by which a fresh liquid-solid interface is continuously created. Near a moving contact line this interface
is yet out of equilibrium and the out-of-equilibrium surface tensions acting on the
contact line ``negotiate'', via  the  Young  equation,  the  value of the contact angle. This contact angle thus depends on the stage of the process, and on the flow, and is, in principle, different from the static
(equilibrium) one. Furthermore, it has been shown that for the capillary rise flow starting from rest the dynamic contact angle can go down from $90$ as the contact line accelerates \citep{SprittlesAndShikhmurzaev2012}. 

The main challenge in the present work consists in fully implementing the interface
formation  model  for  cases  in  which  the  contact  angle  can  vary  in  a  wide  range, in
particular,  for  flows  that  involve  obtuse  contact  angles. The issue at hand is then that, as shown  in  \citet{SprittlesAndShikhmurzaev2011a}, in all models generalising the Navier slip condition, for obtuse
contact angles, a parasitic eigensolution appears, and this  eigensolution gives rise  to  numerical artefacts in the pressure distribution, namely to sharp mesh-dependent spikes with opposite signs near the contact line. An additional difficulty is that the physics of the problem introduces disparate length scales so that accurate numerical implementation of the model, even without addressing the artefact, poses a difficult problem.

In the present work, we address the problem of the implementation of the full interface formation model in the entire range of variation of the dynamic contact angle, in particular addressing the numerical artefact mentioned above, and we show that, for the spreading of a two-dimensional drop from rest over a horizontal solid substrate, there exists a stage of the process where the dynamic contact angle decreases as the contact line accelerates. The structure of the paper is as follows. In  section 2, we summarise  the  interface  formation  equations  and  discuss  the  split-
domain  formulation introduced for its numerical implementation, used to  control the parasitic eigensolution in the  vicinity  of  the contact line. Section 3 discusses the way the split-domain formulation is integrated into the weak form of the equations, and covers important technical aspects of the numerical implementation, including the improvements to previous implementations of the interface formation model. In section 4, the results of our modelling are discussed, and in
section 5 we give a brief review of the implications of these findings and analyse some possible directions of further research.

\section{Problem formulation}\label{sec:Problem_formulation}
We consider the spreading of a 2D drop of an incompressible Newtonian fluid of density $\rho$ and kinematic viscosity $\nu$ over a flat homogeneous solid substrate. If a drop, initially placed on a substrate, is gently pressed against the latter, e.g. by gravity, so as to form a small area of contact, it will begin to spread until the contact angle formed between its free surface and the solid boundary reaches its equilibrium value.

\begin{figure}
\centering
\includegraphics[width=.5\textwidth]{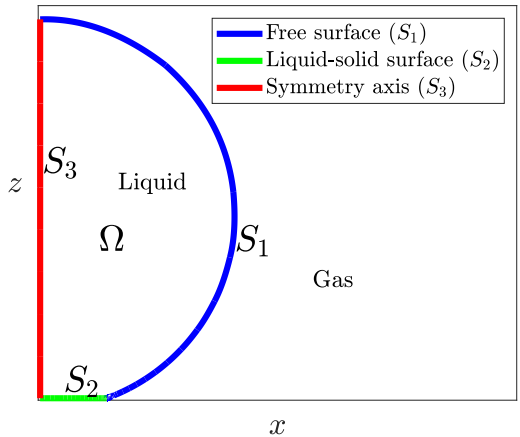}% Here is how to import EPS art
\caption{\label{fig:Domain} Schematics of initial configuration of the droplet. }
\end{figure}

The initial configuration of the droplet $\Omega(t = 0)$ is that of a circular arc with a contact angle of $160^o$ (see the schematic representation in figure \ref{fig:Domain}). The solid occupies the bottom half of the plane and the remaining space is filled by dynamically passive gas with constant pressure $p_g$. The motion of the droplet starts form rest and fluid flow in the droplet is subject to 
gravitational acceleration $\boldsymbol{g}$ and capillary effects. 

We introduce Cartesian coordinates $(x,z)$ with the $x$ axis coinciding with the solid surface and the $z$ axis given by the symmetry line in the 2D droplet, pointing away from the solid. We consider the problem in dimensionless form, using the droplet radius $L \coloneqq R$ as unit length, $U \coloneqq \sigma_{1,e}/\left(\rho\nu\right)$ as unit speed (where $\sigma_{1,e}$ is the equilibrium surface tension of the free surface), $P \coloneqq \rho\nu U/ R$ as the unit stress.

The flow is governed by the Navier-Stokes equation for incompressible flow, i.e.
\begin{equation}\label{eqn:momentum}
    \Rey
    \left[
    \frac{
    \partial \boldsymbol{u}
    }{
    \partial t
    }
    +
    \boldsymbol{u}
    \cdot
    \nabla
    \boldsymbol{u}
    \right] 
    = 
    \nabla
    \cdot
    \mathsfbi{P}
    +
    \St 
    \boldsymbol{g}
\end{equation}
and
\begin{equation}
    \nabla\cdot\boldsymbol{u} = 0,
\end{equation}
where $\boldsymbol{u}$ is the fluid velocity, $\boldsymbol{g} = - g\hat{\boldsymbol{z}}$, $\Rey = \nu U/L^2$ and $\St = gL^2/(\nu U)$, with $U$ and $L$ being the velocity and length scales. Moreover, $\mathsfbi{P}$ is the stress tensor, defined as
\begin{equation}\label{eqn:stress_tensor}
    \mathsfbi{P} = -p\mathsfbi{I}+[\nabla \boldsymbol{u}+(\nabla \boldsymbol{u})^T],
\end{equation}
where $p$ is the pressure and $\mathsfbi{I}$ is the metric tensor.

\subsection{Boundary conditions}
The fluid domain $\Omega(t)$ is bounded by the free surface $S_1$, the liquid-solid surface $S_2$ and the axis of symmetry $S_3$ (figure \ref{fig:Domain}). As required by the interface formation model, we introduce surface velocity and surface density variables $\boldsymbol{v}^s_{i}$ and $\rho^s_{i}$, respectively; which correspond to boundary $S_i$ with $i=1,2$.

\subsubsection{The free surface}
The tangential and normal dynamic boundary conditions (DBC) and the kinematic boundary condition (KBC) are respectively given by
\begin{equation}
    \Ca
    \boldsymbol{n}_1
    \cdot
    \left[
    \nabla
    \boldsymbol{u}
    +
    (
    \nabla 
    \boldsymbol{u}
    )^T
    \right]
    \cdot
    (
    \mathsfbi{I}
    -
    \boldsymbol{n}_1
    \boldsymbol{n}_1
    )
    +
    \nabla
    \sigma_1
    =
    0,
\end{equation}
\begin{equation}
    \Ca\boldsymbol{n}_1\cdot\left\{p_g-p+\boldsymbol{n}_1\cdot\left[\nabla\boldsymbol{u}+(\nabla\boldsymbol{u})^T\right]\right\} = \sigma_1\nabla\cdot\boldsymbol{n}_1
\end{equation}
and
\begin{equation}\label{eqn:KBC_first}
    \left(
    \boldsymbol{c}
    -
    \boldsymbol{v}^s_1
    \right)
    \cdot
    \boldsymbol{n}_1
    =
    0;
\end{equation}
where $\boldsymbol{c} = (u^c,w^c)$ is the velocity of the surface coordinates, $\boldsymbol{n}_1$ is the normal to the free-surface that points into $\Omega$ and $\Ca = \rho\nu U/\sigma_{1,e}$, with $\sigma_{1,e}$ being the equilibrium surface tension coefficient of the free surface, i.e. the surface tension coefficient that is measured when the droplet is static. Following \citet{SprittlesAndShikhmurzaev2012}, the two dynamic boundary conditions can be combined into a single vector boundary condition, given by
\begin{equation}
\label{eqn:DBC1_1}
    \left(
    p_g
    +
    \mathsfbi{P}
    \right)
    \cdot
    \boldsymbol{n}_1
    =
    -
    \frac{
    \nabla^s
    \cdot
    \left[
    \sigma_1
    (
    \mathsfbi{I}
    -
    \boldsymbol{n}_1
    \boldsymbol{n}_1
    )
    \right]
    }{
    \Ca
    }.
\end{equation}

The free-surface and the liquid-solid surface are treated as 2-dimensional phases that exchange mass with the 3-dimensional bulk phase. A surface density is introduced for each interface, which reflects the fact that interfaces are transitional regions with average properties that are different from those of the bulk. An equilibrium surface density exists for each surface, and the deviation of surface tension from its nominal equilibrium value depends (linearly) on how much the surface density deviates from its equilibrium condition. The model also incorporates mass-transport effects along the interfaces and relates surface tension gradients to the slip between bulk and interface phases. Consequently, five further equations are required for the free surface; namely: 
\begin{equation}\label{eqn:SC1}
    \left(
    \mathsfbi{I}
    -
    \boldsymbol{n}_1
    \boldsymbol{n}_1
    \right)
    \cdot
    \left(
    \boldsymbol{v}^s_1
    -
    \boldsymbol{u}
    \right)
    =
    \frac{1+4\bar{\alpha}_g\bar{\beta_g}}{4\bar{\beta_g}}
    \nabla
    \sigma_1,
\end{equation}
\begin{equation}\label{eqn:MEC1}
    \left(
    \boldsymbol{u}
    -
    \boldsymbol{v}^s_1
    \right)
    \cdot
    \boldsymbol{n}_1 
    =
    Q_g
    \left(
    \rho^s_1
    -
    \rho^s_{1,e}
    \right),
\end{equation}
\begin{equation}\label{eqn:TDC1}
    \sigma_1
    =
    \lambda_g
    \left(
    1
    -
    \rho^{s}_1
    \right),
\end{equation}
\begin{equation}\label{eqn:DTC1}
    \epsilon_g
    \left[
    \frac{
    \partial \rho^{s}_1
    }{
    \partial t
    }
    +
    \nabla
    \cdot
    \left(
    \rho^{s}_1
    \boldsymbol{v}^{s}_1
    \right)
    \right]
    =
    \rho^s_{1,e}
    -
    \rho^{s}_1,
\end{equation}
and
\begin{equation}\label{eqn:initial_rhos1}
    \rho^s_1 = \rho^s_{1,e}, \qquad \text{at } t=0;
\end{equation}
where 
$\bar{\alpha}_g = \alpha_g\sigma_{1,e}/(UL)$,
$\bar{\beta}_g = \beta_g UL/\sigma_{1,e}$,
$Q_g = \rho^s_{(0)}/(\rho U\tau_g)$, $\epsilon_g = \tau_g U/L$ and $\lambda_g = \gamma_g \rho^s_{(0)}/\sigma_{1,e}$; with $\alpha_g$, $\sigma_{1,e}$, $\beta_g$, $\rho^s_{(0)}$, $\rho^s_{1,e}$, $\tau_g$ and $\gamma_g$ being physical properties of the surface which are specified in section \ref{sec:Physical_parameters}. 

Equations (\ref{eqn:SC1}-\ref{eqn:TDC1}) respectively correspond to the slip condition between the 1-dimensional (1D) free-surface phase and the bulk phase, the equation for mass exchange between bulk and surface phase, the equation of state that relates surface tension and surface density, and the equation for mass transport along the interface. 

\subsubsection{The liquid-solid surface}
On the liquid-solid surface, the impermeability condition and the generalised Navier-slip condition are respectively given by
\begin{equation}\label{eqn:IC2}
    \left(\boldsymbol{v}^s_2-\boldsymbol{u}_s\right)\cdot\boldsymbol{n}_2 = 0,
\end{equation}
and
\begin{equation}\label{eqn:GNSC}
    {\boldsymbol{n}_2
    \cdot\mathsfbi{P}
    \cdot(
    \mathsfbi{I}
    -
    \boldsymbol{n}_2
    \boldsymbol{n}_2
    ) 
    +
    \frac{
    \nabla\sigma^2}{2\Ca}
    }
    \\
    =
    {
    \bar{\beta}_s
    (
    \boldsymbol{u}
    -
    \boldsymbol{u}_s
    )
    \cdot
    (
    \mathsfbi{I}
    -
    \boldsymbol{n}_2
    \boldsymbol{n}_2
    )
    },   
\end{equation}
where $\boldsymbol{n}_2$ is the normal to the liquid-solid interface that points into $\Omega$ and $\boldsymbol{u}_s$ is the velocity of the solid surface. Here, we include a generic $\boldsymbol{u}_s$ for added generality, despite all cases here considered satisfying $\boldsymbol{u}_s=0$. Moreover,  $\bar{\beta}_s = \beta_s L/(\rho\nu U)$, with $\beta_s$ being a surface property specified in section \ref{sec:Physical_parameters}.

As was the case in the free surface, here we also need further equations, these are
\begin{equation}\label{eqn:SC2}
    \left[
    \boldsymbol{v}^{s}_2
    -
    \frac{1}{2}
    \left(
    \boldsymbol{u}
    +
    \boldsymbol{u}_{s}
    \right)
    \right]
    \cdot
    \left(
    \mathsfbi{I}
    -
    \boldsymbol{n}_2
    \boldsymbol{n}_2
    \right)
    =
    \bar{\alpha}_s
    \nabla
    \sigma_2,
\end{equation}
\begin{equation}\label{eqn:MEC2}
   \left(
    \boldsymbol{u}
    -
    \boldsymbol{v}^{s}_2
    \right)
    \cdot
    \boldsymbol{n}_2
    =
    Q_s
    \left(
    \rho^{s}_2
    -
    \rho^s_{2,e}
    \right),
\end{equation}
\begin{equation}\label{eqn:TDC2}
    \sigma_2
    =
    \lambda_s 
    \left(
    1
    -
    \rho^{s}_2
    \right),
\end{equation}
\begin{equation}\label{eqn:DTC2}
    \epsilon_s
    \left[
    \frac{
    \partial \rho^{s}_2
    }{
    \partial t
    }
    +
    \nabla
    \cdot
    \left(
    \rho^{s}_2
    \boldsymbol{v}^{s}_2
    \right)
    \right]
    =
    \rho^s_{2,e}
    -
    \rho^{s}_2,
\end{equation}
and
\begin{equation}\label{eqn:initial_rhos2}
    \rho^s_2 = 1-\frac{\sigma_{g-s}-\sigma_{1,e}\cos(\theta_c)}{\gamma_s},\qquad \text{at } t = 0;
\end{equation}
where $\bar{\alpha}_s = \alpha_s\sigma_{1,e}/(UL)$, $Q_s = \rho^{s}_{(0)}/(\rho U \tau_s)$, $\epsilon_s = \tau_s U/L$ and $\lambda_s = \gamma_s\rho^{s}_{(0)}/\sigma_{1,e}$;  with $\alpha_s$, $\sigma_{1,e}$, $\rho^s_{(0)}$, $\rho^s_{2,e}$, $\tau_s$ and $\gamma_s$ being surface properties specified in section \ref{sec:Physical_parameters}. The initial condition for equation (\ref{eqn:DTC2}), given by equation (\ref{eqn:initial_rhos2}), is obtained combining equations (\ref{eqn:initial_rhos1}), (\ref{eqn:TDC1}), (\ref{eqn:TDC2}) and the Young equation (see equation \ref{eqn:Young} below).

\subsubsection{The symmetry axis}
On the axis of symmetry the flow is subject to the usual impermeability and no-tangential-stress conditions, given respectively by
\begin{equation}\label{eqn:IC3}
    \boldsymbol{u}\cdot\boldsymbol{n}_3 = 0
\end{equation}
and
\begin{equation}\label{eqn:NTS3}
    \boldsymbol{n}_3
    \cdot
    \mathsfbi{P}
    \cdot
    \left(
    \mathsfbi{I}
    -
    \boldsymbol{n}_3
    \boldsymbol{n}_3
    \right)
    =
    0,
\end{equation}
where $\boldsymbol{n}_3$ is the normal to the axis of symmetry that points into $\Omega$.

\subsubsection{Contact line conditions}
On the free-surface and liquid-solid surface, we have boundary conditions which themselves are partial differential equations (PDEs) that require boundary conditions  of their own. In particular, at the contact line, where these two surfaces meet, they must satisfy Young's equation, namely
\begin{equation}
\label{eqn:Young}
    \sigma_1
    \cos(\theta_c) 
    +
    \sigma_{2}
    =
    \sigma_{g-s},
\end{equation}
where $\sigma_{g-s}$ is the solid-gas surface tension. 

Furthermore, mass flux at the contact line must satisfy
\begin{equation}\label{eqn:MBC}
    \rho^{s}_1
    \left(
    \boldsymbol{v}^{s}_1
    -
    \boldsymbol{U}_c
    \right)
    \cdot
    \boldsymbol{m}_1
    +
    \rho^{s}_2
    \left(
    \boldsymbol{v}^{s}_2
    -
    \boldsymbol{U}_c
    \right)
    \cdot
    \boldsymbol{m}_2
    =
    0,
\end{equation}
where $\boldsymbol{m}_i$, defined only on the boundary of $S_i$, is the unit tangent to boundary $S_i$ that points inward; and $\boldsymbol{U}_c$ is the velocity of the contact line. 

% Moreover, we have
% \begin{equation}
%     \cos(\theta_c)
%     =
%     \boldsymbol{m}^1
%     \cdot
%     \boldsymbol{m}^2.
% \end{equation}

\subsubsection{Conditions at the ends of $S_3$}
At the apex of the droplet, i.e. where the free surface meets the axis of symmetry, we have
\begin{equation}\label{eqn:theta_a}
    \theta_a = \frac{\pi}{2},
\end{equation}
where $\theta_a$ is the angle between the axis of symmetry and the free surface. Moreover, we have
\begin{equation}\label{eqn:ICS1}
    \boldsymbol{v}^s_1\cdot\boldsymbol{n}_3 = 0, \qquad \text{at } x = 0.
\end{equation}
We note that the combination of equations (\ref{eqn:SC1}), (\ref{eqn:IC3}), (\ref{eqn:theta_a}) and (\ref{eqn:ICS1}) implies that, at the apex, we have
\begin{equation}
    \nabla\sigma_1 = 0, \qquad \text{at } x =0,
\end{equation}
since, at this point, $\boldsymbol{n}_3$ is parallel to $\boldsymbol{t}_1$, the unit tangent to the free surface.

Similarly, where the liquid-solid surface meets the axis of symmetry, we also have
\begin{equation}
    \boldsymbol{v}^s_2\cdot\boldsymbol{n}_3 = 0, \qquad \text{at } x =0,
\end{equation}
which, by an argument entirely analogue to the one used above, implies
\begin{equation}\label{eqn:sigma2_origin}
    \nabla\sigma_2 = 0, \qquad \text{at } x = 0.
\end{equation}

Equations (\ref{eqn:momentum})-(\ref{eqn:stress_tensor}), subject to boundary conditions (\ref{eqn:KBC_first})-(\ref{eqn:sigma2_origin}), form the closed system we need to solve. 

\section{Split domain, weak formulation and numerical implementation}\label{sec:Weak_formulation}
We proceed to implement an approximate solution to the system of equations presented above. We first state the equations in an arbitrary Lagrangian-Eulerian reference system, we then introduce a split-domain formulation, which is necessary to solve flows in which the contact angle can be obtuse. Finally, we introduce the weak formulation of the problem and we proceed to apply the finite element method using Galerkin's method. 
\subsection{Arbitrary Lagrangian-Eulerian reference system}
The free-boundary problem at hand is naturally suited for the use of an arbitrary Lagragian-Eulerian (ALE) reference system, which requires that those equations that involve temporal derivatives, i.e. (\ref{eqn:momentum}), (\ref{eqn:DTC1}) and (\ref{eqn:DTC2}), be re-written as
\begin{equation}\label{eqn:momentum_ALE}
    \Rey
    \left[
    \frac{
    \partial \boldsymbol{u}
    }{
    \partial t
    }
    +
    \left(
    \boldsymbol{u}
    -
    \boldsymbol{c}
    \right)
    \cdot 
    \nabla
    \boldsymbol{u}
    \right]
    -
    \nabla
    \cdot
    \mathsfbi{P}
    +
    \St
    \boldsymbol{e}_z 
    =
    0,
\end{equation}
\begin{equation}\label{eqn:DTC1_ALE}
    \epsilon_g
    \left\{
    \frac{
    \partial \rho^{s}_1
    }{
    \partial t
    }
    +
    \rho^{s}_1
    \nabla
    \cdot
    \boldsymbol{c}
    +
    \nabla
    \cdot
    \left[
    \rho^{s}_1
    \left(
    \boldsymbol{v}^{s}_1
    -
    \boldsymbol{c}
    \right)
    \right]
    \right\}
    =
    \rho^s_{1,e}
    -
    \rho^{s}_1,
\end{equation}
and
\begin{equation}\label{eqn:DTC2_ALE}
    \epsilon_s
    \left\{
    \frac{
    \rho^{s}_2
    }{
    \partial t
    }
    +
    \rho^{s}_2
    \nabla
    \cdot
    \boldsymbol{c}
    +
    \nabla
    \cdot
    \left[
    \rho^{s}_2
    \left(
    \boldsymbol{v}^{s}_2
    -
    \boldsymbol{c}
    \right)
    \right]
    \right\}
    =
    \rho^s_{2,e}
    -
    \rho^{s}_2,
\end{equation}
where $\boldsymbol{c}$ is the velocity of the ALE coordinates. We note that, in the latter two equations above, we have used the identity $\nabla^s(f\boldsymbol{h}) = \boldsymbol{h}\cdot\nabla^sf + f\nabla^s\cdot\boldsymbol{h}$, to express the equations in a more convenient format.

It is also worth noting that, in this formulation, we can re-write (\ref{eqn:MBC}) as
\begin{equation}\label{eqn:MBC_ALE}
    \rho^{s}_1
    \left(
    \boldsymbol{v}^{s}_1
    -
    \boldsymbol{c}
    \right)
    \cdot
    \boldsymbol{m}_1
    +
    \rho^{s}_2
    \left(
    \boldsymbol{v}^{s}_2
    -
    \boldsymbol{c}
    \right)
    \cdot
    \boldsymbol{m}_2
    =
    0.
\end{equation}

\subsection{The split-domain formulation}
When the contact angle is acute, the equations listed above can be solved numerically, for instance, using the finite element method (FEM), as done in \citet{SprittlesAndShikhmurzaev2013}. However, as was shown by \citet{SprittlesAndShikhmurzaev2011b}, when the contact angle is obtuse, a straightforward computational implementation of this set of equations leads to nonphysical solutions that dominate the flow around the contact line. These effects are indeed observed with any of the standard FEM implementations for dynamic wetting, regardless of the inclusion of the interface formation equations. 

In \cite{SprittlesAndShikhmurzaev2011a}, the origin of such nonphysical behaviour is found, and a solution to the problem is provided. The undesired behaviour is caused by the numerical method picking up a member of a one-dimensional space of solutions to the problem of Stokes' flow in a wedge (subject to no-tangential-stress conditions on its boundaries). The removal of these nonphysical features is achieved by expressing the velocity field, in the vicinity of the contact line, as a sum of one of this eigensolution (whose amplitude is to be determined), and a second unknown component of the velocity field. Effectively, this means that we remove the eigensolution contribution when calculating the remaining components of the flow. The eigensolution contribution is later put back, completing the correct physical picture.

We can thus write
\begin{equation}\label{eqn:u_plus_ucheck}
    \boldsymbol{u} = \bar{\boldsymbol{u}} + A(t)\check{\boldsymbol{u}},
\end{equation}
where $\check{\boldsymbol{u}}$ is the solution to the problem
\begin{equation}
    \nabla\cdot\mathsfbi{P} = 0,
\end{equation}
\begin{equation}
    \nabla\cdot\check{\boldsymbol{u}} = 0 
\end{equation}
on the plane, which is subject to
\begin{equation}
    \boldsymbol{n}
    \cdot
    \mathsfbi{P}
    \cdot
    \left(
    \mathsfbi{I}
    -
    \boldsymbol{n}
    \boldsymbol{n}
    \right)
    =
    0,
\end{equation}
along the lines than bound a wedge of angle $\theta_c$.

The conditions above determine the eigensolution to be given by the stream function
\begin{equation}
    \psi = B\left(\check{x}^2+\check{z}^2\right)^{\lambda/2}
    \sin 
    \left(\lambda\left[
    \theta
    +\theta_c
    -\pi\right]\right),
\end{equation}
where $\lambda = \pi/\theta_c$, $B$ is an arbitrary factor (to be set as $B=1$), $\check{x} = x - x_c$, and $\check{z} = z - z_c$, with $(x_c,z_c)$ being the coordinates of the contact line. We note that this is a solution on the entire plane, not just the wedge, though boundary conditions are satisfied at the boundary of the wedge. This fact is important, as it allows us to evaluate the solution slightly outside the wedge, which can be necessary in this implementation.

A factor $A(t)$, the ``amplitude'' of the eigensolution, multiplied by $\psi$, becomes one more unknown in the system, and an additional equation is required to determine it, namely
\begin{equation}\label{eqn:p_limit}
    \lim_{r\to 0}\frac{\partial p}{\partial \theta} = 0,
\end{equation}
where $r = \sqrt{\check{x}^2+\check{z}^2}$,  $\theta = \tan^{-1}\left(-\check{z}/\check{x}\right)$, with $0 < \theta < \pi$.

\begin{figure}
\centering
\includegraphics[width=.5\textwidth]{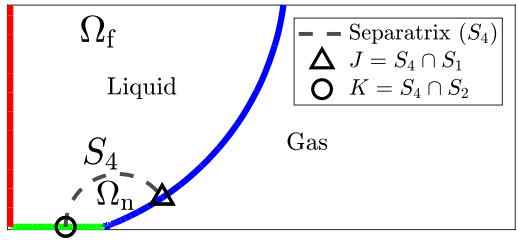}% Here is how to import EPS art
\caption{\label{fig:Split_domain} Schematic representation of the split domain. Compatibility conditions for the bulk equations are applied along the $S_4$ line, and continuity is assumed for the interface formation variables on points $J$ and $K$.}
\end{figure}

The introduction of the eigensolution is only necessary in the vicinity of the contact line, and it is therefore convenient to split the $\Omega$ domain into two sub-domains, $\Omega_{\text{n}}$ and $\Omega_{\text{f}}$, respectively spanning the near-field of the contact line and the far-field (see figure \ref{fig:Split_domain}). 

In this split-domain formulation we solve two sets of PDEs; which, at the separatrix of the respective domains ($S_4$), are subject to the following compatibility conditions 
\begin{equation}
    \boldsymbol{u}_{\text{f}} = \bar{\boldsymbol{u}} + A\check{\boldsymbol{u}},
\end{equation}
where $\boldsymbol{u}_{\text{f}}$ is the velocity as calculated in the far field; and
% \begin{equation}
%     p_{\text{f}} = p_{\text{n}},
% \end{equation}
\begin{equation}
    \mathsfbi{P}_{\text{f}}\cdot\boldsymbol{n}_4 
    =
    \mathsfbi{P}_{\text{n}}\cdot\boldsymbol{n}_4, 
\end{equation}
where $\boldsymbol{n}_4$ points into the far field domain. 

Equations for flow in the vicinity of an obtuse contact angle are obtained by substituting (\ref{eqn:u_plus_ucheck}), into the formulation given by (\ref{eqn:momentum})-(\ref{eqn:MBC}). The equations on $\Omega_{\text{f}}$ and $\Omega_{\text{n}}$ involve functions that are defined in both sub-domains, such as $p$, $\boldsymbol{v}^s_i$, $\rho^s_i$ and $\sigma_i$, for $i =1,2$. Naturally, these must be continuous across $S_4$.  

It should be emphasised that the split-domain formulation is only necessary while the contact angle is obtuse. However, in this problem (as well as in many other dynamic wetting ones), the contact angle is free to fluctuate between acute and obtuse; consequently, it is more convenient to always act as if we are solving two sets of PDEs in their respective domains; the far field problem being fixed and the near field problem being switched from the obtuse-contact-angle formulation to the acute one, simply by forcing $A$ to be zero and removing equation (\ref{eqn:p_limit}) when needed.

% the boundary conditions are also are required to satisfy the following compatibility conditions
% \begin{equation}
%     \boldsymbol{u}^s_{i,\text{f}} 
%     =
%     \boldsymbol{u}^s_{i,\text{n}},
% \end{equation}
% \begin{equation}
%     \rho^s_{i,\text{f}} 
%     = 
%     \rho^s_{i,\text{n}},
% \end{equation}
% and 
% \begin{equation}
%     \sigma_{i,\text{f}} 
%     =
%     \sigma_{i,\text{n}},
% \end{equation}
% at $S_i\cap S_4$, with $i = 1,2$, and the sub-indices $\text{f}$ and $\text{n}$ indicating on which domain the variables are defined.

\subsection{Weak formulation}
Once the split-domain problem is clearly defined, we introduce its weak formulation, in preparation for the use of the finite element method.

The split domain formulation re-shapes our original bulk PDEs in the following form
\begin{eqnarray}
    \chi_{\Omega_{\text{f}}}
    \mathcal{P}_{\text{f}}
    \left(
    \boldsymbol{u},
    p,
    \boldsymbol{u}^s_1,
    \rho^s_1,
    \sigma_1,
    \boldsymbol{u}^s_2,
    \rho^s_2,
    \sigma_2
    \right)
    \ \ \ \ \ \ \ \ \ \ \nonumber
    \\
    +
    \chi_{\Omega_{\text{n}}}
    \mathcal{P}_{\text{n}}
    \left(
    \bar{\boldsymbol{u}},
    p,
    \boldsymbol{u}^s_1,
    \rho^s_1,
    \sigma_1,
    \boldsymbol{u}^s_2,
    \rho^s_2,
    \sigma_2
    \right)
    = 0,
\end{eqnarray}
where $\mathcal{P}_{\text{f}}$ and $\mathcal{P}_{\text{n}}$ represent the differential operators in the far-field and near-field bulk, respectively. Consequently, the weak form can be written as
\begin{equation}\label{eqn:Weak_Split}
    \int
    \limits_{\Omega_{\text{f}}}
    {
    \phi_i
    \mathcal{P}_{\text{f}}
    }
    +
    \int
    \limits_{\Omega_{\text{n}}}
    {
    \phi_i
    \mathcal{P}_{\text{n}}
    }
    =
    0,
\end{equation}
where $\phi_i$ is an arbitrary element of our set of test functions and $\Omega_{\text{f}}$ and $\Omega_{\text{n}}$ are the far-field and near-field sub-domains, respectively.

It is important to highlight that the PDEs being solved on both sub-domains are the same. That is, we are still solving a smooth PDE in the entire domain. It is only the expression of the bulk velocity that changes, and the form in which we write the equations changes as a consequence of this.

Similarly, for each of the surface PDEs in the interface formation model we have
\begin{eqnarray}
    \chi_{S_{j,\text{f}}}
    \mathcal{P}^j_{\text{f}}
    \left(
    \boldsymbol{u},
    p,
    \boldsymbol{u}^s_j,
    \rho^s_j,
    \sigma_j
    \right)
    \ \ \ \ \ \ \ \ \ \ \nonumber
    \\
    +
    \chi_{S_{j,\text{n}}}
    \mathcal{P}^j_{\text{n}}
    \left(
    \bar{\boldsymbol{u}},
    p,
    \boldsymbol{u}^s_j,
    \rho^s_j,
    \sigma_j
    \right)
    = 0,
\end{eqnarray}
with $j = 1,2$. For a given element $\phi^j_i$ of our set of test functions on the boundary, we have
\begin{equation}\label{eqn:Weak_Split_bound}
    \int
    \limits_{S_{j,\text{f}}}
    {
    \phi^j_i
    \mathcal{P}^j_{\text{f}}
    }
    +
    \int
    \limits_{S_{j,\text{n}}}
    {
    \phi^j_i
    \mathcal{P}^j_{\text{n}}
    }
    =
    0.
\end{equation}

The form above implies that residuals associated to test functions that are supported on the union of the near-field and far-field sub-domains will, in general, include a contribution from each. The expressions for each residual contribution in the far and near fields are given in Appendix \ref{app:Residuals}.

\subsection{Numerical Implementation}\label{sec:Numerical_implementation}
We use Galerkin's method with the FEM on the weak form of the split-domain formulation of the problem in an Arbitrary Lagrangian-Eulerian system of reference. We follow closely the methodology of \citet{SprittlesAndShikhmurzaev2012}, including their choice to use the spine method, the V6P3 taylor-Hood triangular elements, and their meshing algorithm based on the bi-polar system of coordinates, where the separatrix between near-field and far-field sub-domains happens along one of the circular spines. However, the exact features of our mesh are somewhat different, so as to optimise it for our applications.

Close to the contact line, the spines are separated using the criteria given by \citet{SprittlesAndShikhmurzaev2012}, i.e. the distance between consecutive spines along the horizontal liquid-solid line increases in a geometric sequence that uses the ratio recommended in \citet{SprittlesAndShikhmurzaev2012} (ratio = 1.07). Next to this region, we use a very similar geometric criteria, this time with a ratio of 1.1. We also introduce a geometric progression for the separation between the spines along the horizontal axis in the vicinity of the axis of symmetry, that is, we define an initial separation for the spines closest to the vertical axis, and we place neighbouring spines with increasing separation in a geometric sequence of ratio $1.2$. The remaining middle section of the liquid-solid line is filled with equally spaced spines, the spacing in this middle section is chosen as close as possible to the average of the two spacings to its left and right. As the flow progresses, the length of the liquid-solid line (which is considered a spine in the problem) changes, and the separation between points of intersection of the circular spines with the horizontal axis change in proportion.

\begin{figure}
    \centering
    \includegraphics[width=.8\textwidth]{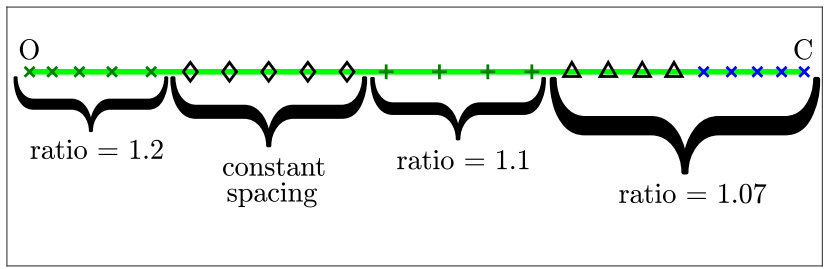}
    \caption{Schematic representation of the spine spacing along the horizontal axis. Bullets correspond to intersection between the spines and the liquid-solid line, point $O$ corresponds to the axis of symmetry and point $C$ to the contact line.}
    \label{fig:Spine_feet_spacing}
\end{figure}

Given the location of the contact line, the separation between the axis of symmetry and its neighbouring spine, and the separation between the contact line and its neighbouring circular spine; all we need to place every spine is the number of spines in each of the sections where the geometric progression for the spacing applies. The number of spines in the intermediate section will be determined by the size of the spacing which is chosen as described above. See figure \ref{fig:Spine_feet_spacing} for an schematic representation.

Another modification to the meshing algorithm presented on \citet{SprittlesAndShikhmurzaev2012}
 is necessary in order to verify equation (\ref{eqn:p_limit}). The numerical implementation of this limit is done by ensuring that the two pressure nodes on each interface (one on $S_1$ and one on $S_2$) that are at the closest (strictly positive) distance to the contact line, are located on a circumference centred at the contact line. This is solved assigning the centre of the circular spine with the two smallest radii to be given the contact line itself. When the mesh is chosen in this way, the numerical verification of equation (\ref{eqn:p_limit}) is simply done by equating the pressure at these two nodes, as is recommended in \citet{SprittlesAndShikhmurzaev2011a}.

The intersection of each circular spine with the horizontal axis is enough to define the radius of each spine (level set of the bipolar coordinate system), as explained in \citet{SprittlesAndShikhmurzaev2012}; since this is determined by the initial distribution and the length of the liquid-solid line, the length of all spines (including the liquid-solid line) is all that is needed to fully determine the shape of the domain. 

\begin{figure}
    \centering
    \includegraphics[width = .5\textwidth]{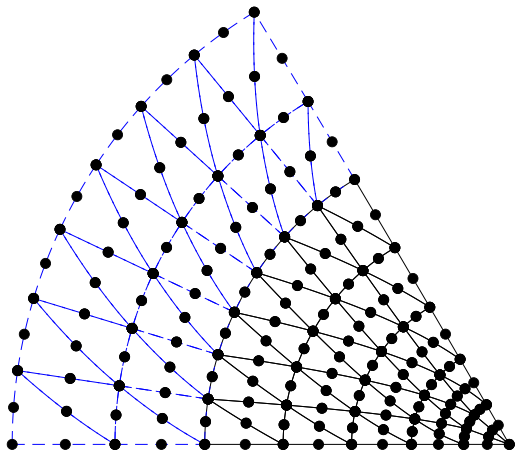}
    \caption{Detail of the mesh in the vicinity of the contact line.}
    \label{fig:near_field_mesh}
\end{figure}

Spines are numbered from $1$ to $n_s$, with the first spine being the liquid-solid line and the subsequent ones listed in increasing order of distance to the contact line. Spine $n_s$ is on the axis of symmetry.

The mesh in \citet{SprittlesAndShikhmurzaev2012} had a single element to have a vertex on the contact line. This was perfectly justified in the cases there considered, as they only ever dealt contact angles that are less than or equal to $90^o$. However, in our case, when the angle can be close to $180^o$, this convention would deform this element almost to the point of degeneracy. We thus allow a generic number of elements at the contact line, three being sufficient for all purposes here considered.

As in \citet{SprittlesAndShikhmurzaev2012}, our elements are placed along circular bands around the contact line,as can be seen in figure \ref{fig:near_field_mesh}. Each band has nodes along three spines.  As shown in figure \ref{fig:near_field_mesh}, our mesh has three triangular elements in touch with the contact line the contact line. The number of elements placed along the following band of elements is of seven and each subsequent band includes two elements more than the number of elements in the prior band. This pattern is maintained over the first portion of the region where we use the factor of $1.07$ to determine the progression of the spacing between spines (see the $\times$ markers next to the contact line in figure \ref{fig:Spine_feet_spacing}). The first band formed by the remaining spines in this first section ($\triangle$ markers) includes a single element more that the prior band, and all subsequent bands in the entire mesh keep the exact same number of elements as the prior band, as shown in dashed blue lines in figure \ref{fig:near_field_mesh}. 

The separatrix between near and far field is given by the last solid black spine in figure \ref{fig:near_field_mesh}, and in practice we use $38$ circular spines with the $1.07$ spacing ratio, with $14$ of those in the region next to the contact line, $115$ spines in the region with the factor of $1.1$, and 21 in the region with the factor of $1.2$. Moreover, for the configuration shown in the sections below, we use a minimum length that $2^{12}$ times smaller than the minimum recommended by \citet{SprittlesAndShikhmurzaev2013}.

Gaussian integration is used to numerically find the residuals and their derivatives. We use an 8th order rule with 16 points based on permutation stars on a triangle, following \citet{ZhangEtAl2009}. The exact eigensolution is sampled at the Gaussian integration points, as opposed to it being sampled at the nodes and interpolated by the hat functions, following the advise given in \cite{SprittlesAndShikhmurzaev2011a}. To capture the singular nature of the pressure at the contact line, we follow \cite{SprittlesAndShikhmurzaev2011b} and use an interpolating function for pressure that is singular at that point. 

Temporal derivatives are discretised using the implicit, second-order, variable-step-size, backward differentiation formula (VS-BDF2) following \citet{CelayaEtAl2014} and \citet{Denner2014}. The initial time step is obtained using the implicit Euler method. The first time step size is taken to be the characteristic time divided by a power of $2$, which is chosen to guarantee that the change in angle is less than one degree in this step. Subsequent time-steps are set fixing a maximum and minimum desired change in the contact angle. If the prior time step violated either condition, the next time step is chosen as double or half the prior one, accordingly. In the case discussed in the sections below we used a maximum change in dynamic contact angle of $1^o$ and a minimum change in contact angle of $0.1^o$, over the first portion of the simulation; and when the Newton method started taking more than five iterations to converge, we changed this numbers to $0.2^o$ and $0.02^o$, respectively. To verify whether our choice of time-step is adequate, we run the same simulation with half the chosen time-step until it reaches maximum velocity. We then compare the resulting predictions for the dynamic contact angle, and we accept the simulation if they differ by less than $0.1^o$.

Five improvements, with respect to what was done in \citet{SprittlesAndShikhmurzaev2013} and \citet{SprittlesAndShikhmurzaev2012PoF}, to the weak form of the interface formation residuals are introduced here. First, equations are re-cast so that the surface tension functions do not strictly need to be differentiable; secondly, certain terms in the momentum equations are exchanged for boundary terms using Gauss' theorem and the incompressibility condition (reducing the computations needed); thirdly, residuals for equations (\ref{eqn:DTC1_ALE}) and (\ref{eqn:DTC2_ALE}) are expressed in way that does not require the free-surface normal to be differentiated (thus removing all need for the calculation of second derivatives to find the residuals). More importantly, the fourth improvement relates to the implementation of equation (\ref{eqn:MBC_ALE}), which is done in a way that is symmetric and more robust. Finally, the fifth improvement consists in equations (\ref{eqn:MEC1}) and (\ref{eqn:MEC2}) being combined with (\ref{eqn:DTC1}) and (\ref{eqn:DTC2}) to produce more numerically tractable residuals. The residual equations used in this work are given in full detail in appendix \ref{app:Residuals}, where each of these improvements is shown.

Out of these improvements, we highlight the symmetric implementation of equation (\ref{eqn:MBC_ALE}). In prior implementations, this equation was used exclusively to feed information from the contact line into the liquid-solid surface \citep[see equation 43]{SprittlesAndShikhmurzaev2013}; whereas, in the present work, information can flow from one boundary to the other as dictated by the flow.

The eigensolution is switched off (i.e. $A \equiv 0$) for contact angles lower than $\pi/2$, which can happen halfway through a Newton method iteration, and therefore the standard Newton algothim has to be adapted to select the correct system of equations to solve.

It is extremely important to highlight that when using the eigensolution in the near field, the angle that one should use to calculate variable $\lambda$ is the angle measured on the mesh $\theta_{c,m}$, i.e. the one given by the solid surface and the tangent to the free-surface line element at the contact line. This angle converges in the Newton method iterations to the value of $\theta_c$, i.e. the variable that appears in the momentum and other equations, but if $\theta_c$ is used to calculate $\lambda$ at the start of the Newton method iterations, the method shows a tendency to escape convergence. Convergence of these two angles is absolutely vital for a consistent simulation. In fact, it is the criterion used in \citet{SprittlesAndShikhmurzaev2012} to identify a sufficiently refined mesh, which was also used here.

Care should be taken when calculating this angle and any other quantity that depends on the tangents to the free surface. In particular, in the vicinity of the contact line, elements are extremely small, so it is often convenient to translate their coordinates to a point in the element before operating on them. Moreover, when finding lengths in these elements, small quantities are squared, then added, and the result is raised to the $1/2$ power; hence, it is often convenient to normalise the numbers by a typical size of them before squaring them. One can then find the lengths and, finally, re-scale them.

We note that the advice for the minimum length that needs to be resolved, given in \citet{SprittlesAndShikhmurzaev2013}, appears to be sufficient only for acute angles. The larger the angle, the more this length needs to be reduced. Here we reduce the length by trial and error, until the result of the simulations shows that the contact angle, as measured in from the mesh, and the dynamic contact angle variable $\theta_c$ are within $0.2^o$ of each other.

All entries in the Jacobian of the system of equations are calculated analytically, with the only exception of the entries in the column that correspond to the derivatives with respect to the length of the liquid-solid line; for these entries, derivatives are calculated using the chain rule, with the derivative of the coordinates of the nodes found numerically, while every other derivative in the chain is found analytically. 

All code was written in Matlab and is made available at
\newline \href{https://github.com/grcarlosa/IFM\_repo}{https://github.com/grcarlosa/IFM\_repo}. This includes the meshing algorithms, construction of the residuals and the Jacobian of the system, as well as the implementation of the modified Newton method. Extensive notes on these derivations are available at the public repository in \href{https://github.com/grcarlosa/IFM\_notes\_repo}{https://github.com/grcarlosa/IFM\_notes\_repo}.

\subsection{Physical constants}\label{sec:Physical_parameters}
Bulk fluid constants are taken from standard values for water at $25^o$C, i.e. 
\begin{equation}
    \rho = 1\times 10^{3}\, \rm{kg/m}^3
\end{equation}
\begin{equation}
    \nu = 1\times 10 ^{-6}\, \rm{m}^2/\rm{s} 
\end{equation}
and an equilibrium surface tension for the free surface of
\begin{equation}\label{eqn:sigma1_e_dim}
    \sigma_{1,e} = 7.286\times 10 ^{-2}\, N/m.
\end{equation}
Gravitational effects are not considered, i.e. we take
$g = 0\, \rm{m/s}^2$,
and we assume a constant atmospheric pressure of the gas phase, which we take as $p_g = 0\, \rm{N}/\rm{m}^2$.

The characteristic units are given by $L = R = 0.5\,\mu$m , $U = \sigma_{1,e}/\left(\rho\nu\right) = 72.86$\,m$/$s and $P = \rho\nu U/ R = 145,720\,$ N$/$m$^2$. Interface formation constants are taken following the recommendations of \citet{SprittlesAndShikhmurzaev2013}, \citet{SprittlesAndShikhmurzaev2012PoF}. In particular, we assumed a width of the interfacial layer of 
\begin{equation}
    \upsilon = 1\times 10^{-9}\rm{m},
\end{equation}
also following \citep{BazhlekovAndChesters1996numerical}, and the relaxation times for the interfaces are given by
\begin{equation}
    \tau_s
    =
    7\times 10^{-6}\frac{\rm{m}\,\rm{s}^2}{\rm{kg}}
    =
    7\times 10 ^{-12}\, \rm{s},
\end{equation}
with $\tau_g = \tau_s/100$.

From the assumptions above, it follows that
\begin{equation}\label{eqn:rhos0}
    \rho^s_{(0)} 
    = 
    \rho \upsilon %    =    (1\times 10^{3}\, \rm{kg/m}^3)    (1\times 10^{-9}\rm{m})
    =
    1\times 10^{-6}\, \rm{kg}/\rm{m}^2,
\end{equation}
\begin{equation}
    \alpha_g 
    =
    \alpha_s
    =
    \frac{\upsilon}{\rho \nu}
    =%\frac{1\times 10^{-9}\rm{m}}{(1\times 10^{3}\rm{kg}/\rm{m}^3)\,(1\times 10 ^{-6}\, \rm{m}^2/\rm{s}),}=
    1\times 10^{-6}\,\rm{m}^2\rm{s}/kg,
\end{equation}
\begin{equation}
    \beta_g
    =
    \beta_s
    =
    \frac{\rho\nu}{\upsilon}
    =%\frac{(1\times 10^{3}\rm{kg}/\rm{m}^3)(1\times 10 ^{-6}\, \rm{m}^2/\rm{s})}{1\times 10^{-9}\rm{m}}=
    1\times 10^{6}\, \rm{kg}/(\rm{m}^2\rm{s}),
\end{equation}
and we take
\begin{equation}
    \rho^s_{1,e}
    = 0.6 \rho^s_{(0)}
    =
    6\times 10^{-7}\, \rm{kg}/\rm{m}^2,
\end{equation}
which, together with equations (\ref{eqn:TDC1}), (\ref{eqn:sigma1_e_dim}) and (\ref{eqn:rhos0}), implies
\begin{equation}
    \gamma_g
    =
    \frac{\sigma_{1,e}}{\rho^s_{(0)}-\rho^s_{1,e}}
    =%    \frac{(72.86\times 10 ^{-3}\, N/m)}{(1\times 10^{-6}-6\times 10^{-7})\,\rm{kg}/\rm{m}^2}=
    1.82\times 10^{5}\, \rm{m}^2/\rm{s}^2,
\end{equation}
and we take $\gamma_s = \gamma_g$.

We assume an equilibrium contact angle of
\begin{equation}
    \theta_{c,e} = \pi/3,
\end{equation}
and a gas-solid surface tension $\sigma_{g-s} = 0$,
which from the Young equation (\ref{eqn:Young}) and (\ref{eqn:sigma1_e_dim}) imply
\begin{equation}
    \sigma_{2,e} = -3.64\times 10 ^{-2}\, N/m.
\end{equation}
In turn, from equation (\ref{eqn:TDC2}), it follows that
\begin{equation}
    \rho^s_{2,e}
    =
    1.2\times 10^{-6}\, \rm{kg}/\rm{m}^2.
\end{equation} 

We these physical parameters, we obtain the following dimensionless numbers $\Rey = 36.43$, $\St = 0$, $\Ca = 1$, $\bar{\alpha}_g = 0.002$, $\bar{\beta}_g = 5\times 10^{5}$, $\epsilon_g = 0.0102$, $L_g = Q_g \epsilon_g = 0.002$ (see equation \ref{eqn:E1_re-arranged}), $\bar{\lambda}_g = 2.5$, $\bar{\beta}_s = 500$, $\bar{\alpha}_s = 0.002$, $\epsilon_s = 1.02$, $L_s = Q_s \epsilon_s = 0.002$, $\bar{\gamma}_s = 2.5$.
 
\section{Results}\label{sec:Results}
Simulations show an initial acceleration of the contact line, followed a comparatively slow descend of the contact-line velocity. The contact angle is monotonically decreasing from start to end of the contact-line motion. Consequently, we observe a decreasing contact angle at an accelerating contact line over the initial portion of the simulation, followed by a longer regime in which the contact line velocity drops to zero as the contact angle continues to relax to its equilibrium value.% (see figure \ref{fig:Ucl_and_theta_c_plus_zoom}).

\begin{figure}
    \centering
    \begin{tabular}{cc}
    \includegraphics*[width = .4\textwidth]{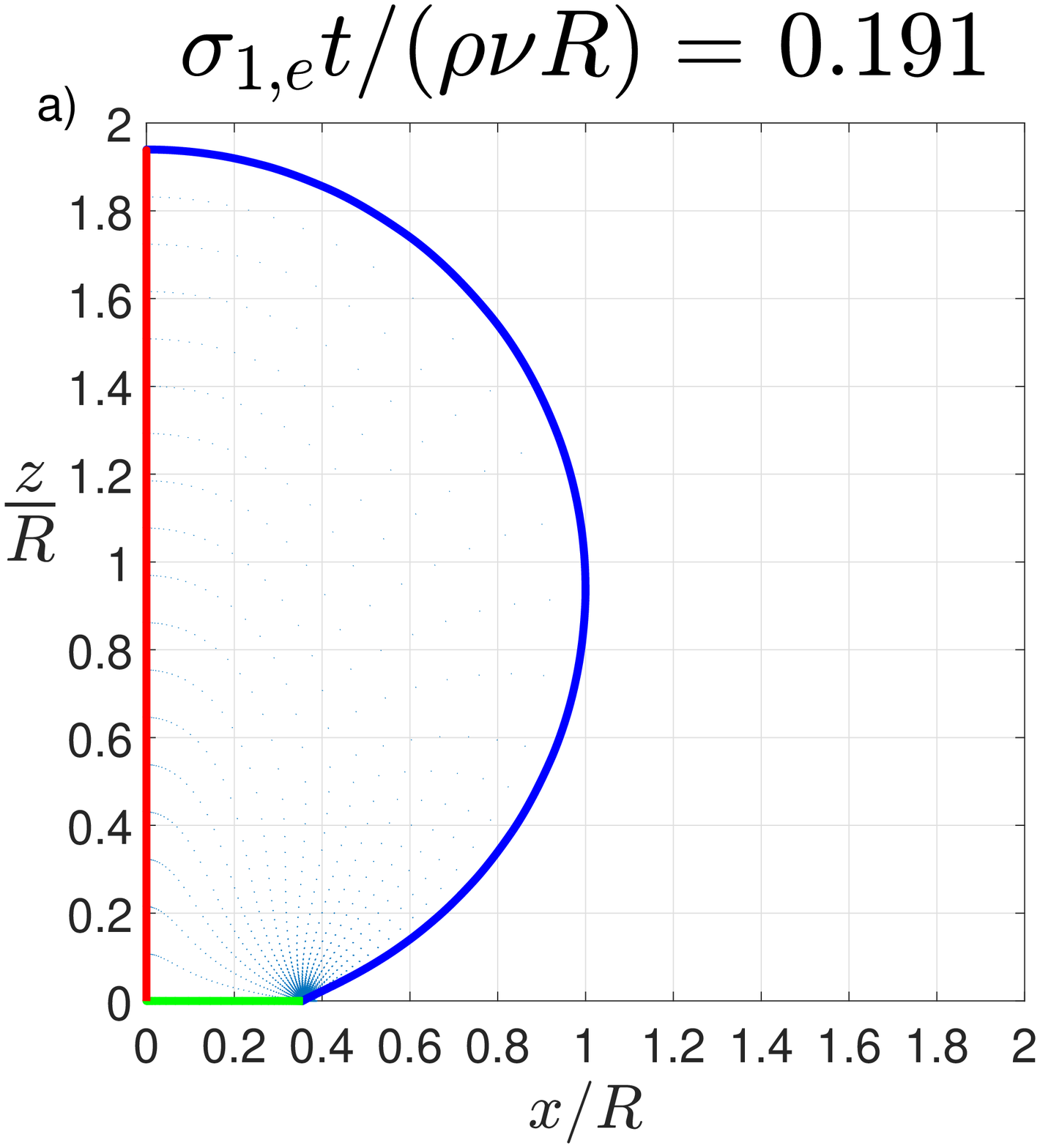}
    &  
    \includegraphics*[width = .4\textwidth]{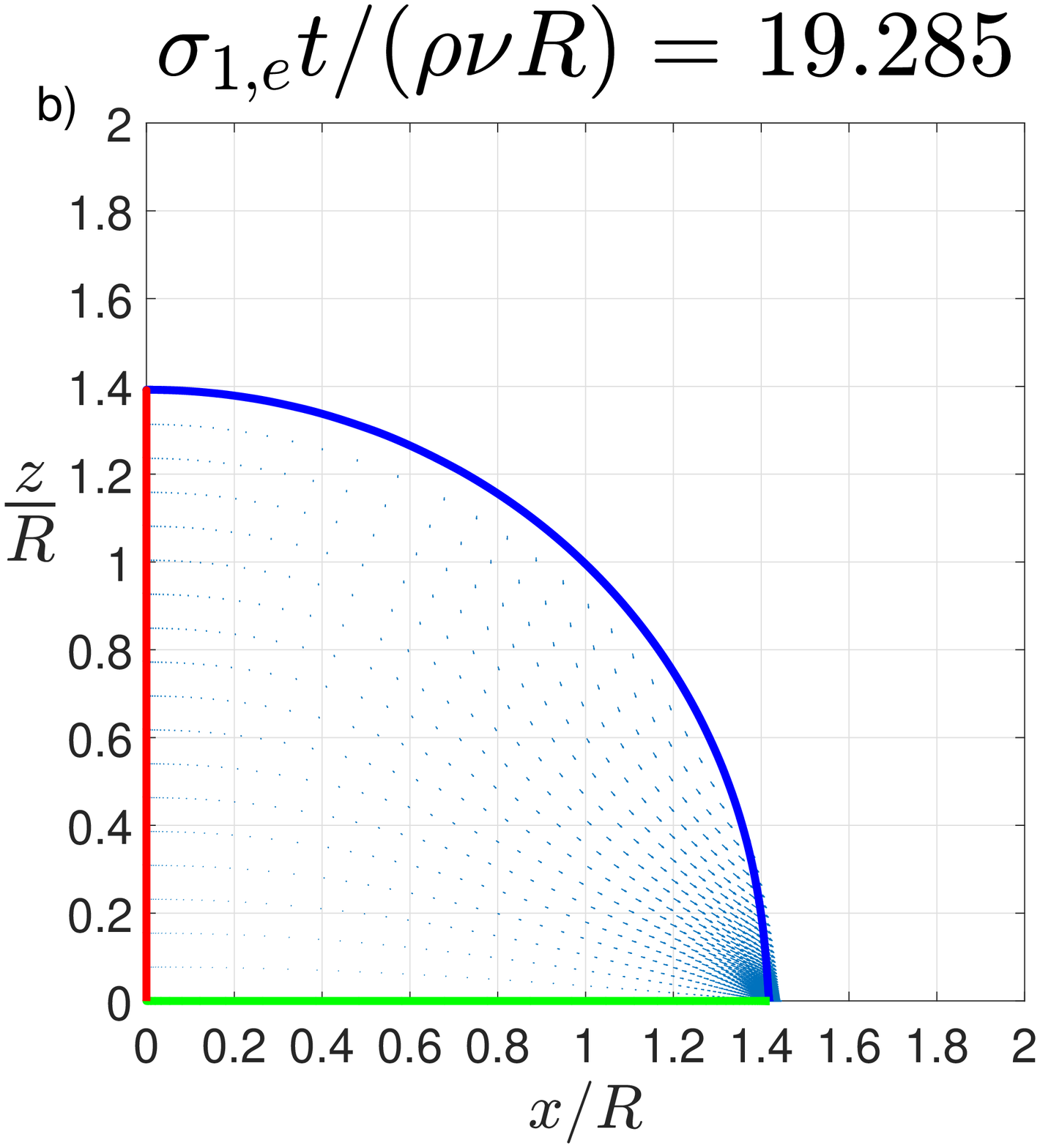}
    \\
    \includegraphics*[width = .4\textwidth]{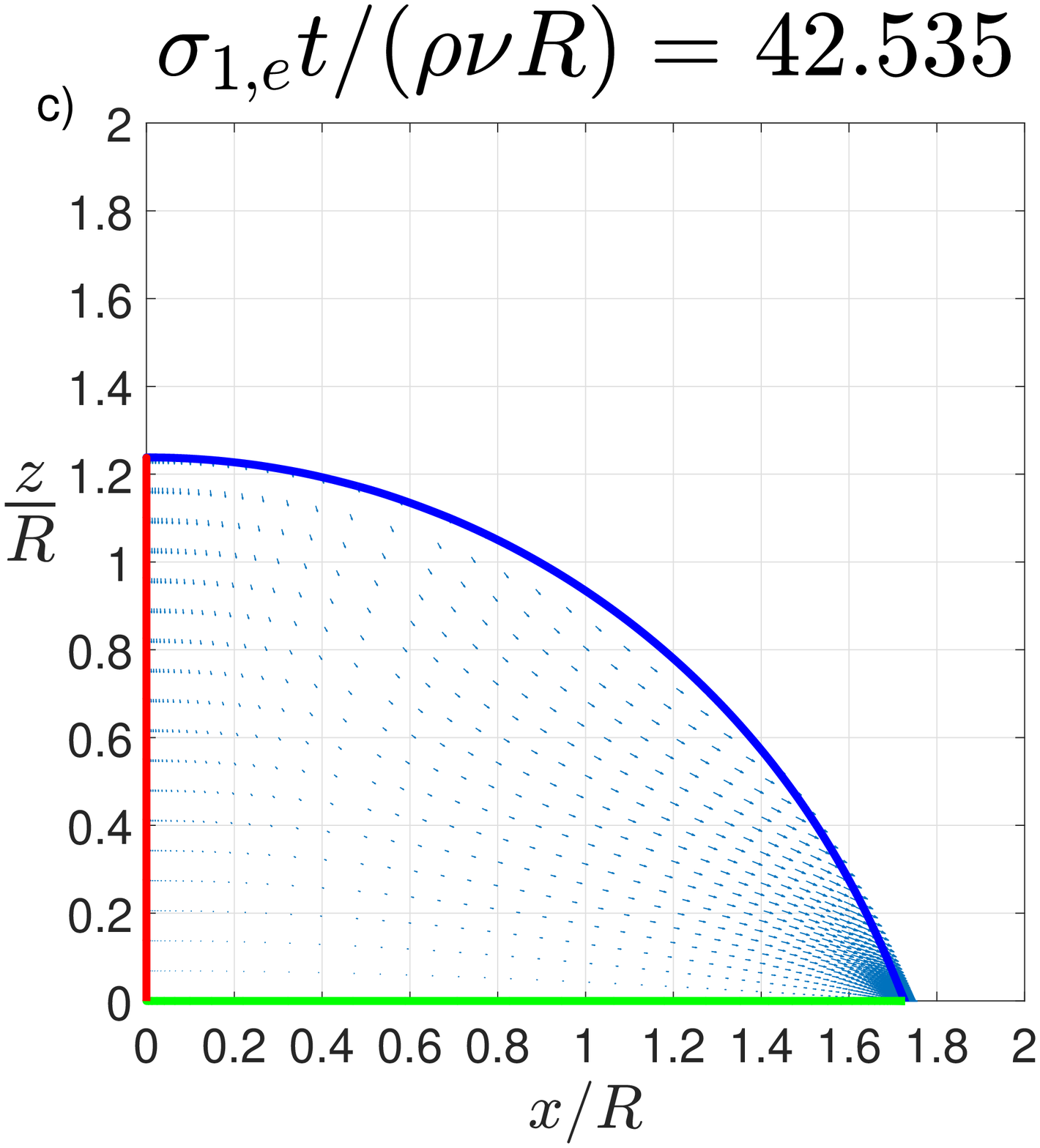}
    &  
    \includegraphics*[width = .4\textwidth]{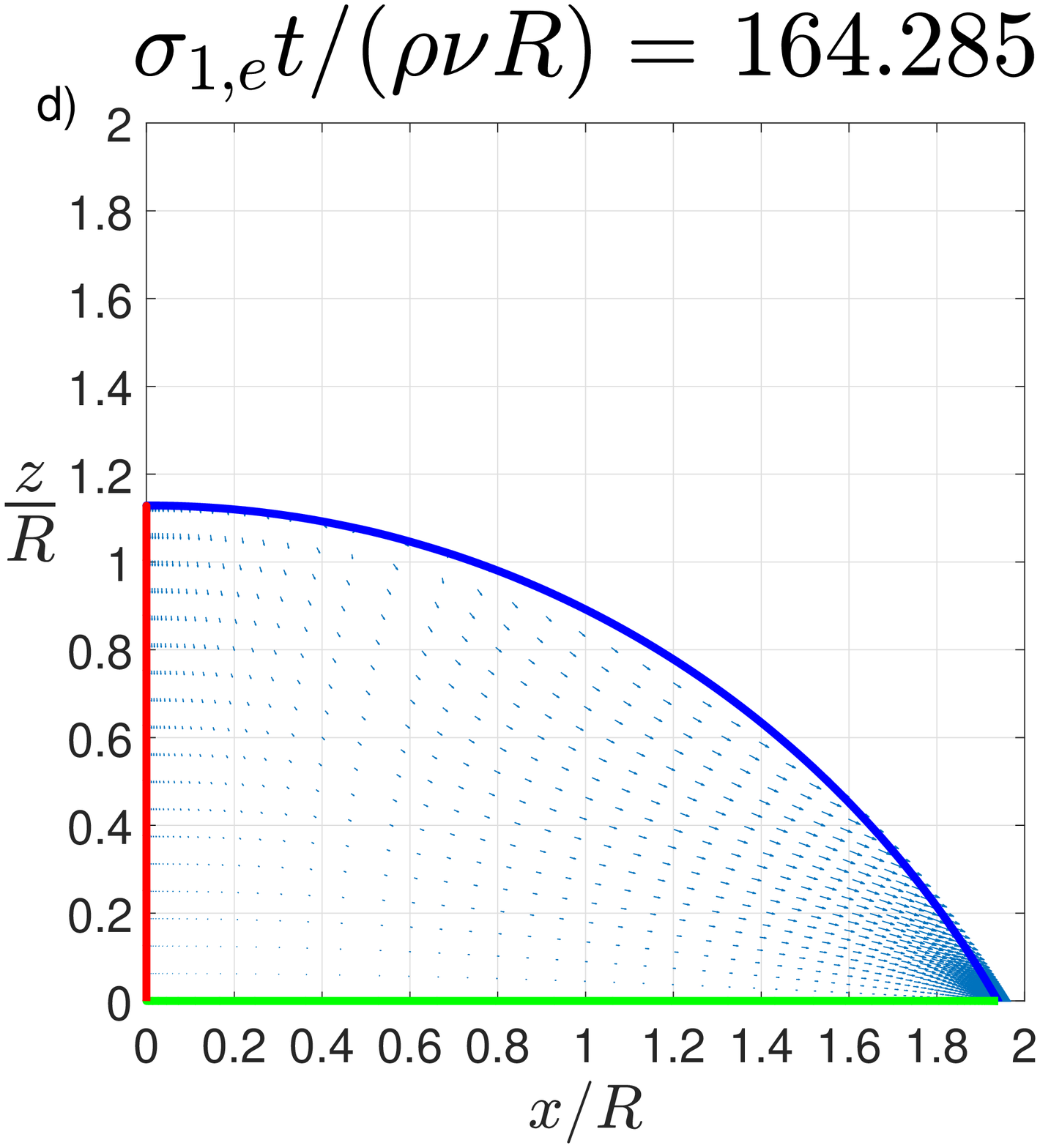}
    \\
    \end{tabular}
    \caption{Large-scale view of the spreading process. Arrows show the flow velocity in the frame of reference of the stationary solid and their lengths are re-scaled at each time to better illustrate relative distribution of the flow at each time.}
    \label{fig:spread}
\end{figure}
Motion of the droplet is initiated as the out-of-equilibrium surface density along the liquid-solid line drives mass exchange with the bulk and mass transport along the surface, driving the liquid-solid surface density towards its equilibrium. The free surface is sucked into the liquid-solid surface, thus perturbing the free-surface density in the vicinity of the contact line, away from its initial equilibrium. These changes in surface densities imply changes in surface tension, which control the contact angle and induce Marangoni flow in the bulk, resulting in a fully capillary-driven spontaneous flow of the droplet towards its equilibrium configuration.

\begin{figure}
    \centering
    \begin{tabular}{cc}
    \includegraphics*[width = .4\textwidth]{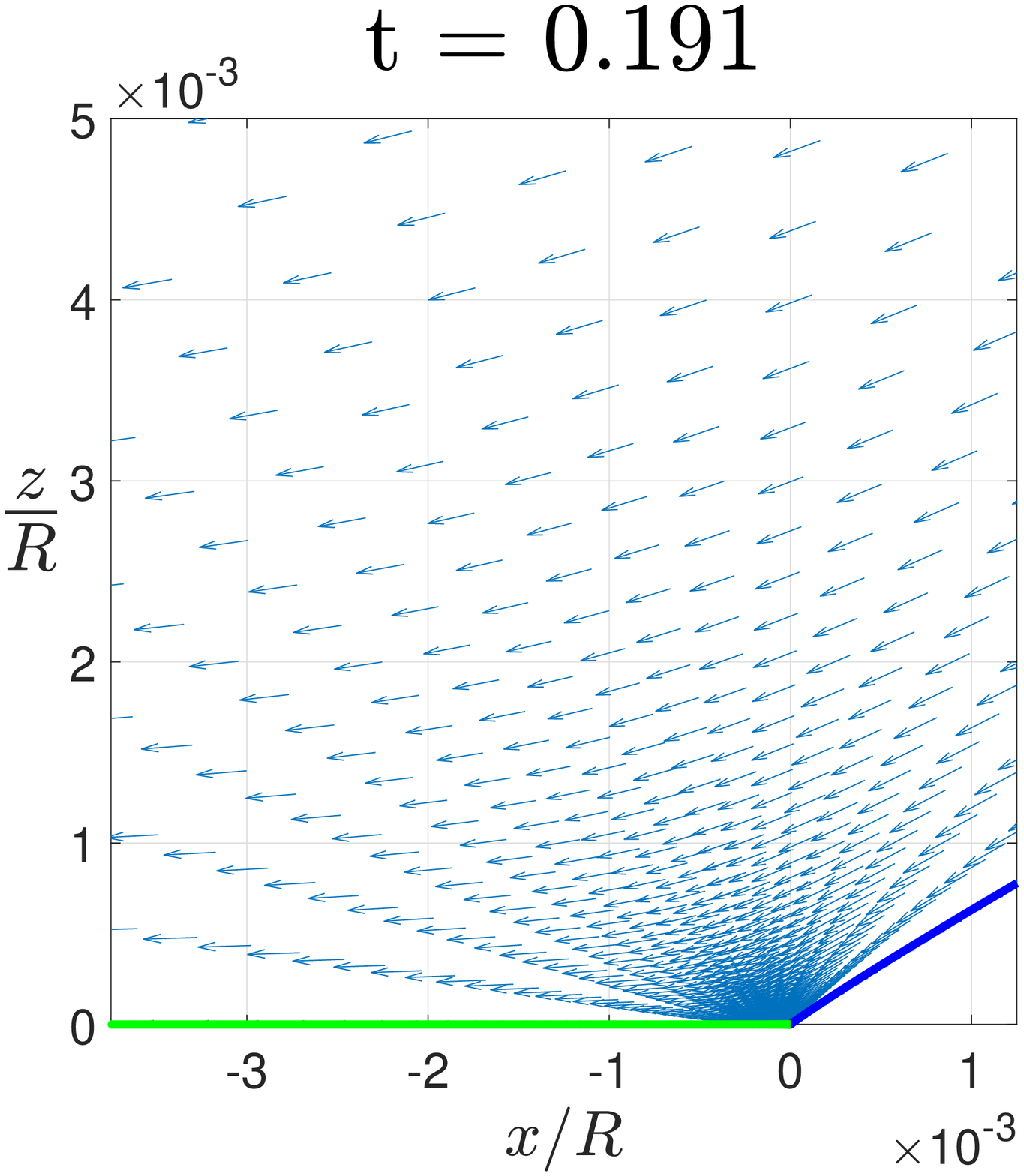}
    &  
    \includegraphics*[width = .4\textwidth]{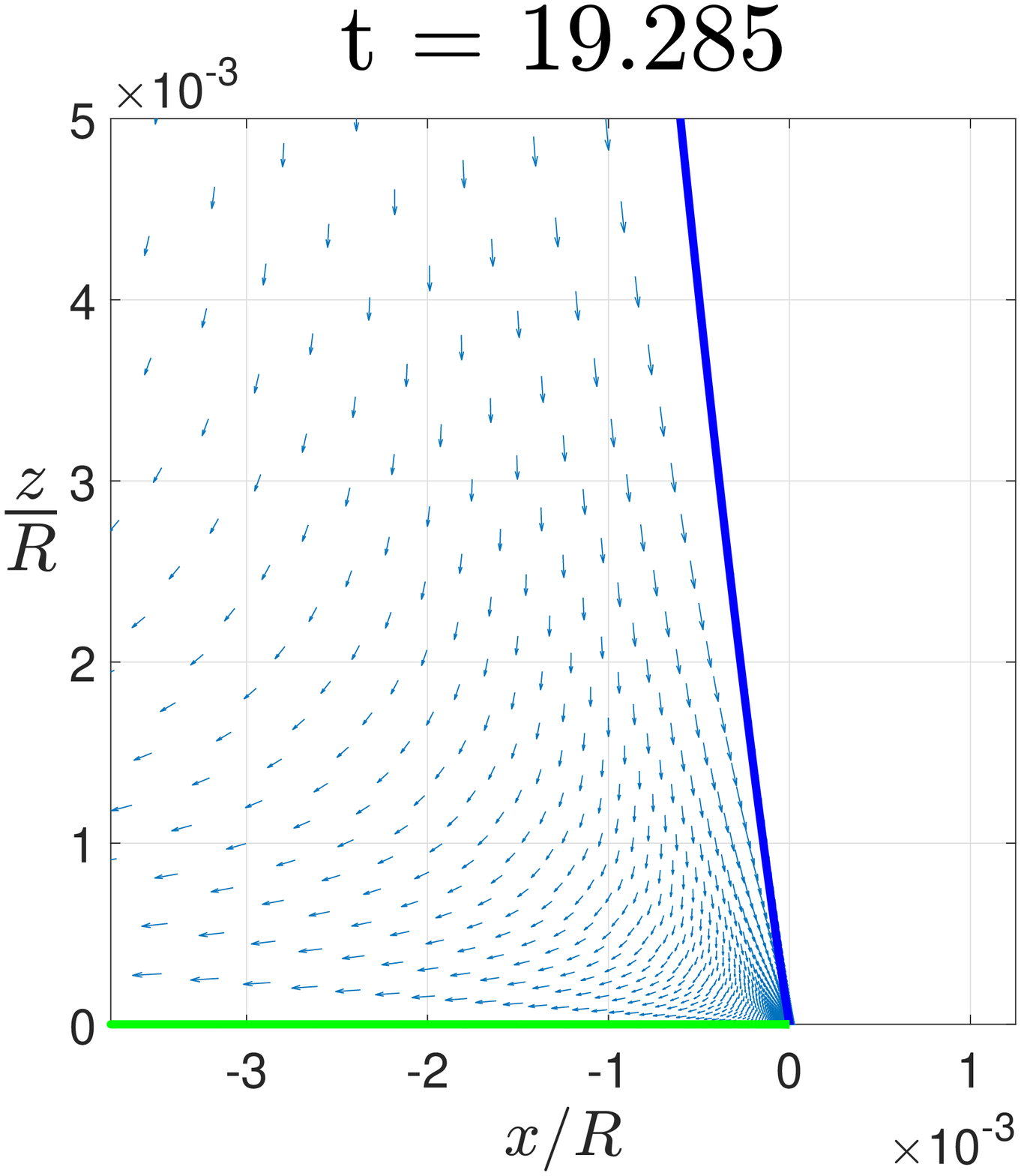}
    \\
    \includegraphics*[width = .4\textwidth]{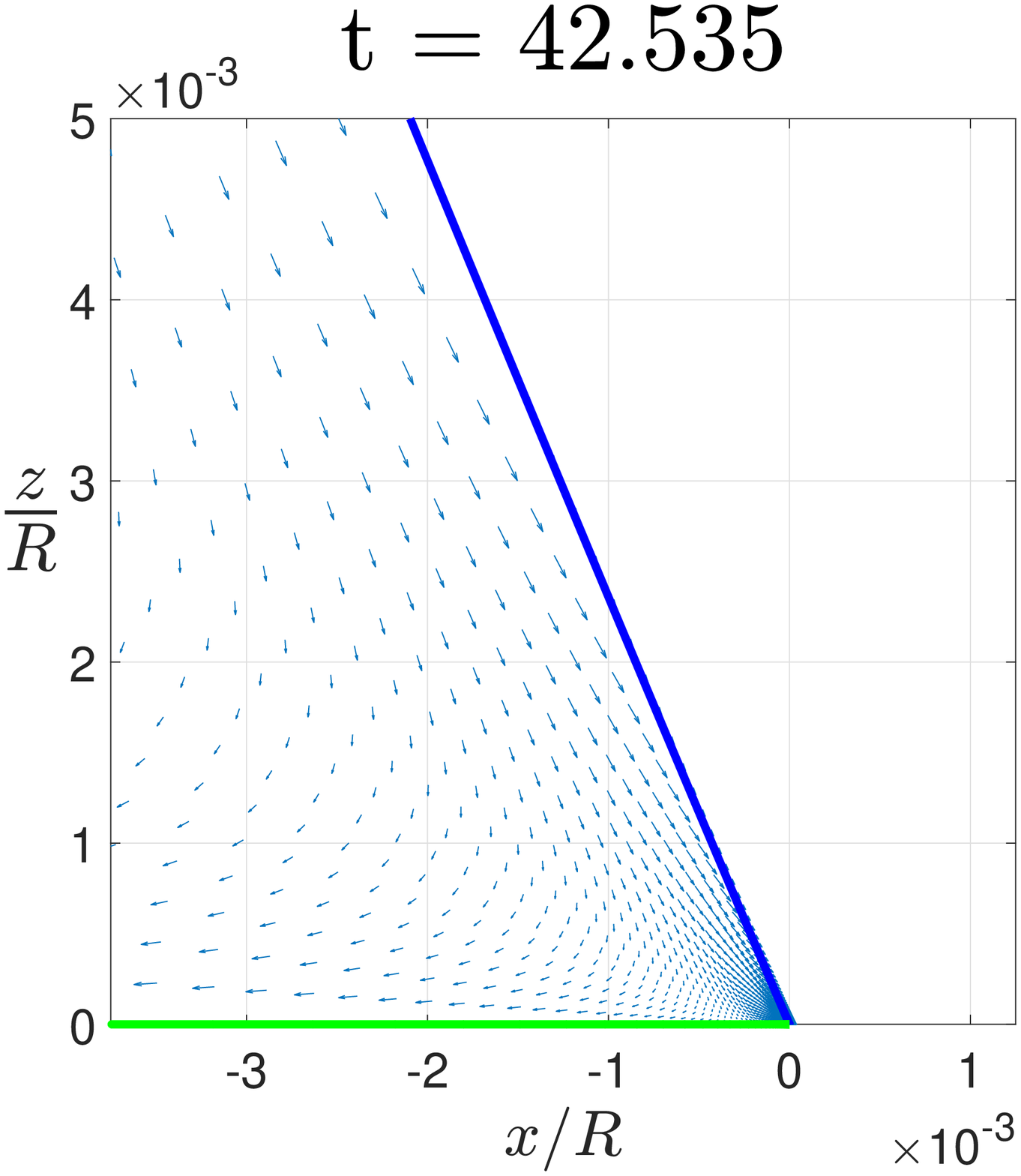}
    &  
    \includegraphics*[width = .4\textwidth]{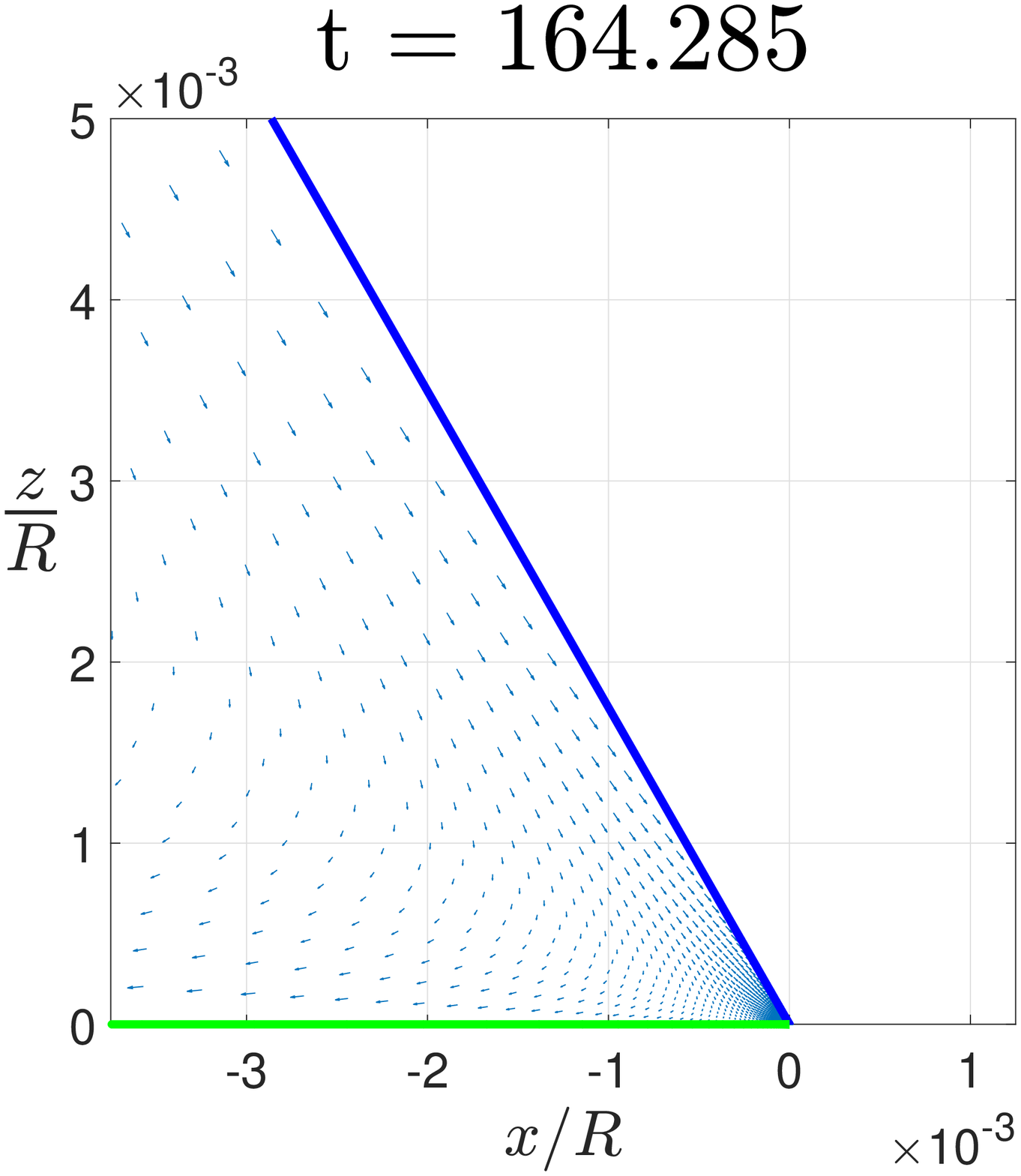}
    \\
    \end{tabular}
    \caption{Close-up of the velocity fields in the frame of reference of the moving contact line. Arrow lengths are re-scaled at each time to better illustrate spatial distribution of velocity.}
    \label{fig:spread_zoom}
\end{figure}

Flow in the bulk is initially concentrated near the contact line (see figure \ref{fig:spread}), where we do not observe the noisy behaviour associated with the artefact described in \citet{SprittlesAndShikhmurzaev2011a}, indicating that the removal of such artefacts, by means of the use of our split-domain formulation and eigensolution, was successful (see figure \ref{fig:spread_zoom}). A video animation of the flow field is provided as supplementary material.

\begin{figure}
    \centering
    \begin{tabular}{lr}
    \includegraphics[width = .435\textwidth]{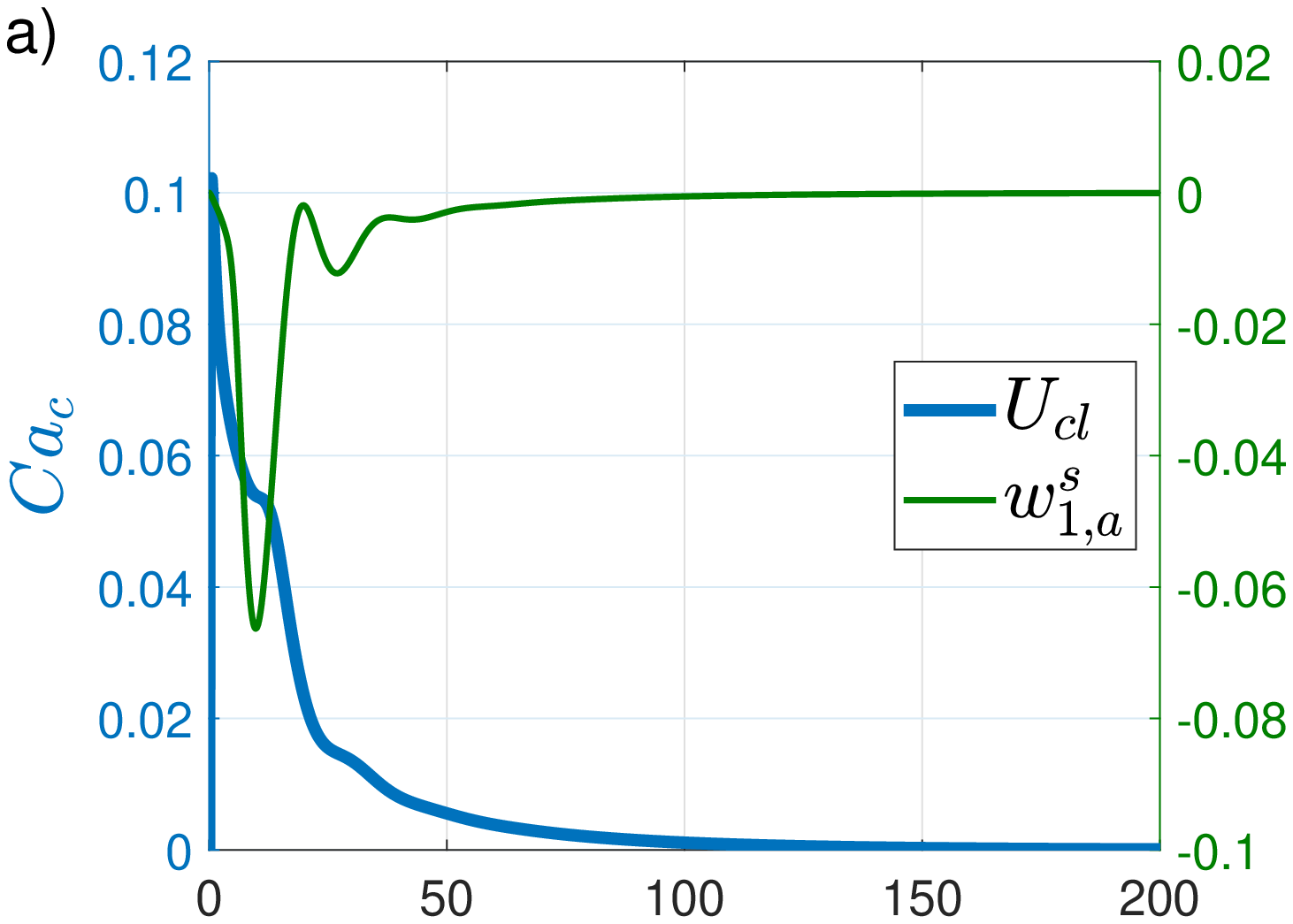}
    &  
    \includegraphics[width = .421\textwidth]{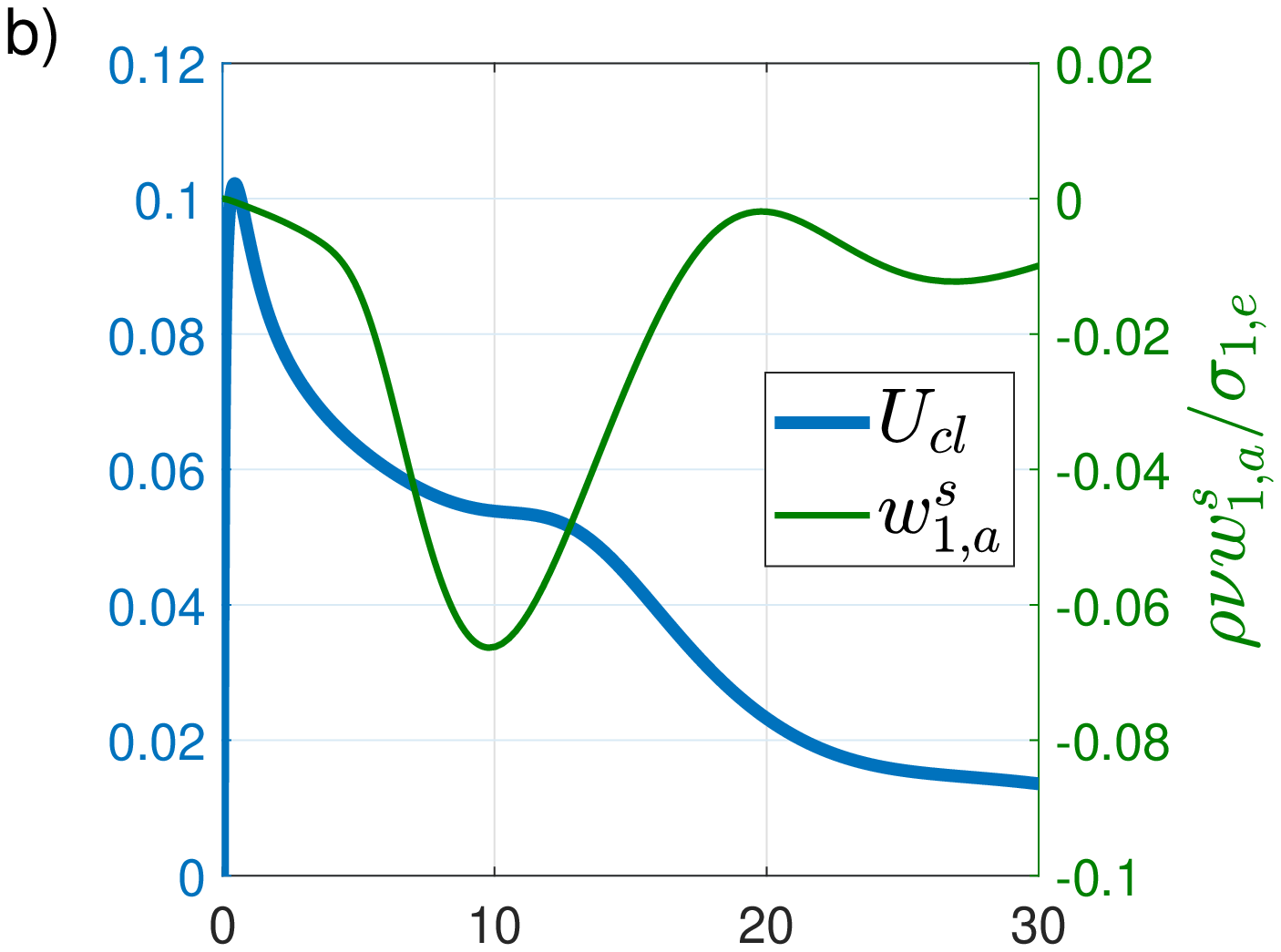}
    \hspace*{2mm}
    \\
    \vspace*{1.5mm}
    \includegraphics[width = .44\textwidth]{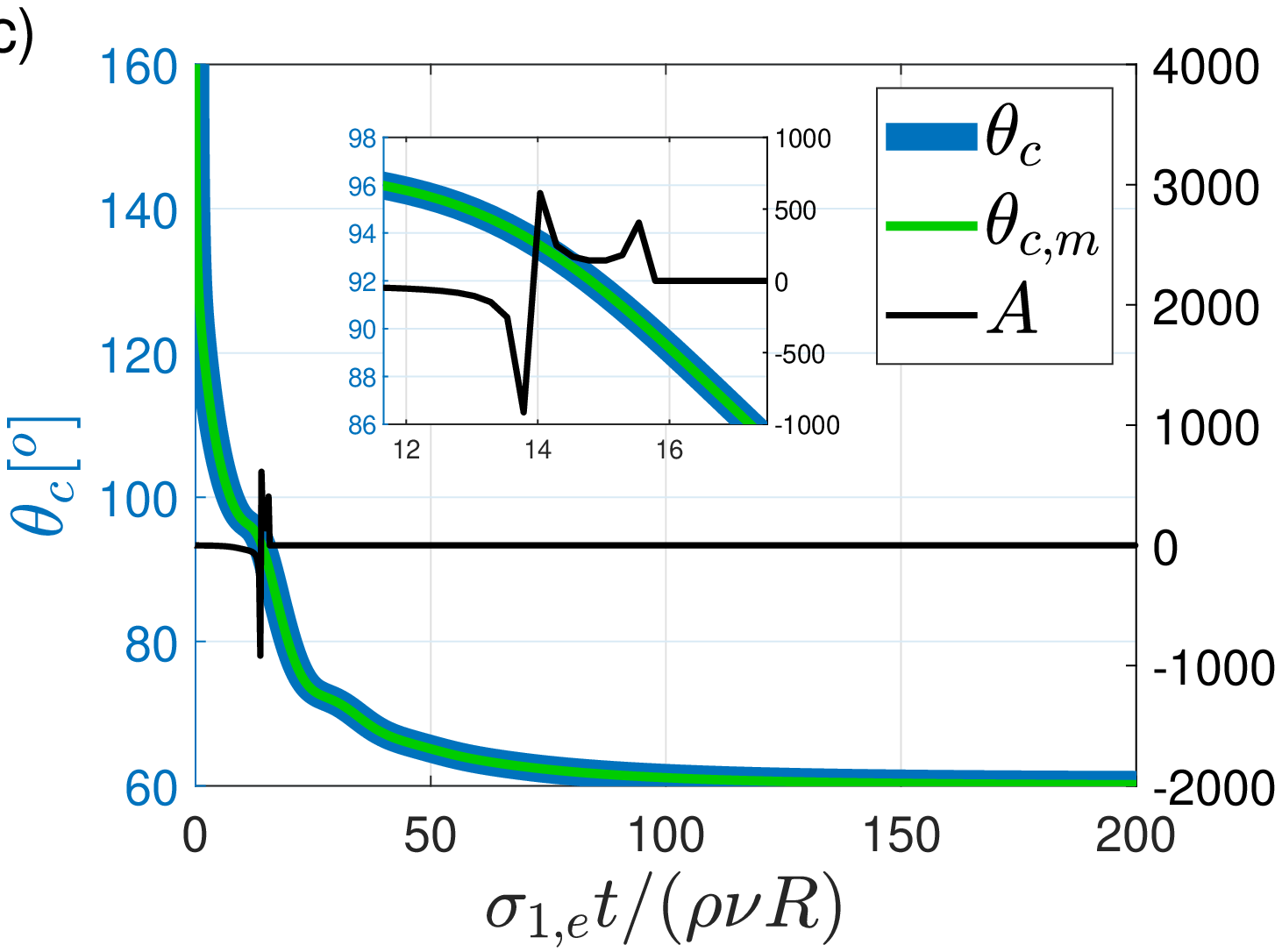}
    &  
    \includegraphics[width = .43\textwidth]{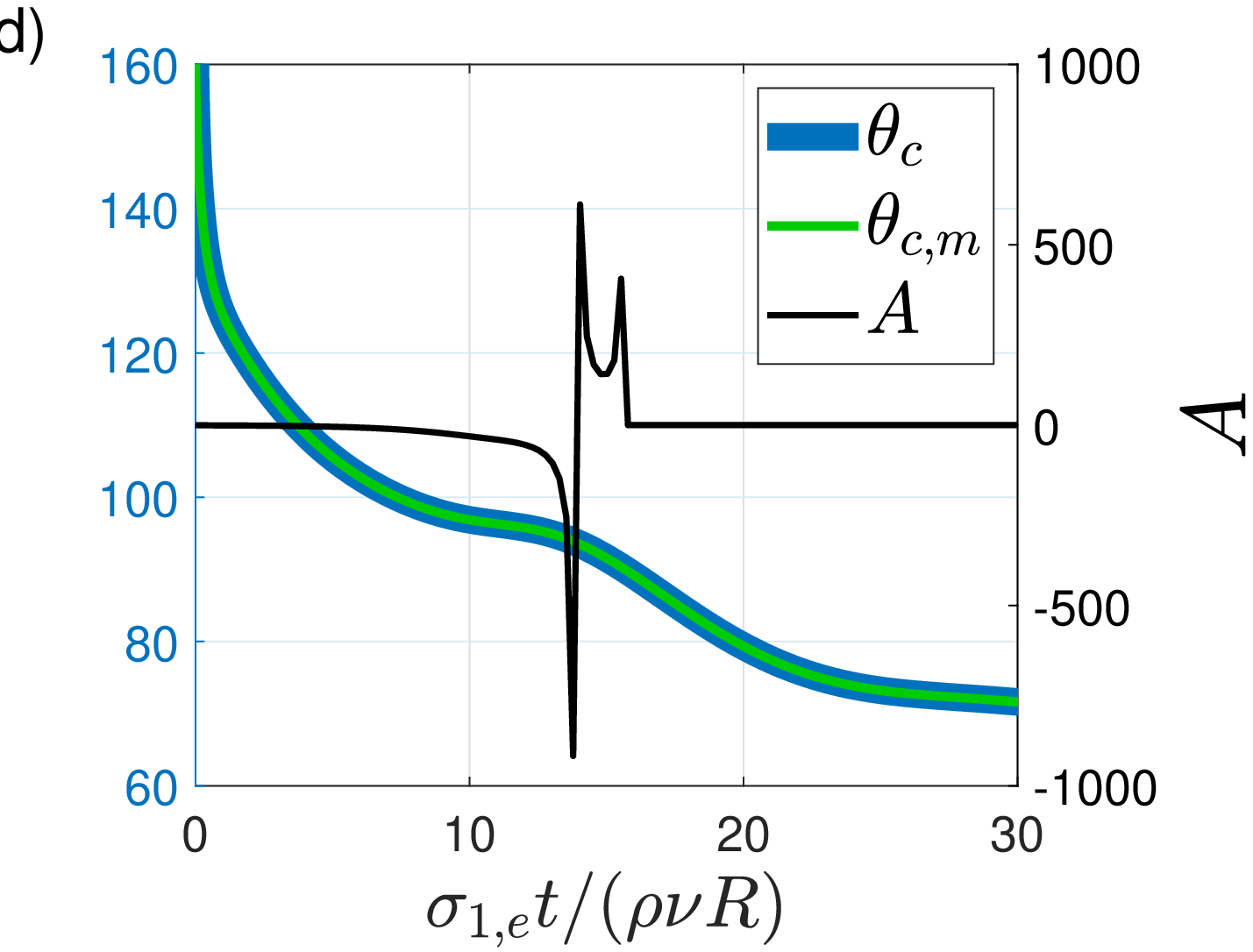}
    \hspace{.5mm}
    \end{tabular}
    \caption{Contact line velocity and contact angle as functions of time. Panels correspond to the microsecond-scale process for velocity (a) and contact angle (c), and the nanosecond-scale regime of contact line acceleration (b) and initial drop in the contact angle (d).}
    \label{fig:Ucl_and_theta_c_plus_zoom}
\end{figure}

Figure \ref{fig:Ucl_and_theta_c_plus_zoom} shows the evolution of the contact-line velocity and contact angle over the full process (panel a for velocity and c for contact angle), as well as a detail of the early stages (panel b for velocity and d for contact angle). Panels a and b also show the vertical velocity of the free surface at the apex of the droplet. Similarly, panels c and d also compare the dynamic contact angle (i.e. the result of the solution of the system of equations for variable $\theta_c$) with the contact angle measured from the tangent to the free surface at the contact line $\theta_{c,m}$. The latter angle is ultimately determined by the length of the two circular spines that are closest to the contact line. Moreover, panels c and d, also show the amplitude of the eigensolution contribution needed to satisfy the limit condition of axial symmetry of pressure as one approaches the contact line. 

Oscillations in the contact line velocity and contact angle curve in figures \ref{fig:Ucl_and_theta_c_plus_zoom}a and \ref{fig:Ucl_and_theta_c_plus_zoom}c, correspond to the apex of the droplet suddenly descending faster and slowing down soon after that. This reflects the ability of the model to capture effects that the global configuration of the flow has on the contact line behaviour.

As can be seen in figures \ref{fig:Ucl_and_theta_c_plus_zoom}c and \ref{fig:Ucl_and_theta_c_plus_zoom}d, the two measurements of the contact angle ($\theta_c$ and $\theta_{c,m}$) are virtually indistinguishable. The convergence of these two ways of measuring the contact angle was used as the criterion to determine that our mesh was sufficiently refined, as recommended by \citet{SprittlesAndShikhmurzaev2013}.

Figure \ref{fig:Ucl_and_theta_c_plus_zoom}c illustrates how massively important the contribution of the eigensolution becomes for angles slightly above $90^o$, in agreement with what was reported in \citet{SprittlesAndShikhmurzaev2011a}. The eigensolution contribution is present from the very start of the simulation, though not with amplitudes comparable to those found when the contact angle is slightly beyond $90^o$.

\begin{figure}
    \centering
    \includegraphics[width=\textwidth]{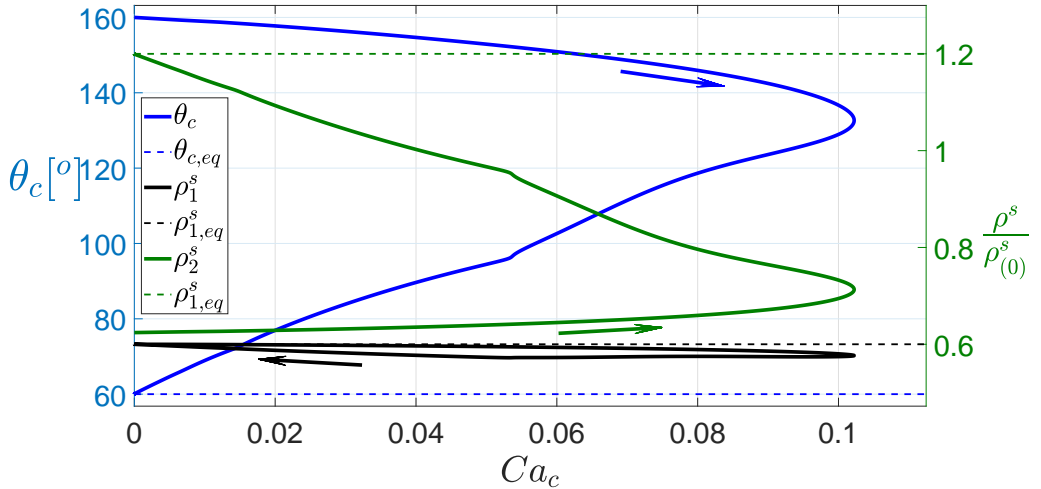}
    \caption{Process trajectories for contact angle and surface densities at the contact line.}
    \label{fig:Summary_CL}
\end{figure}

The resulting relation between contact line velocity and contact angle is multi-valued, as predicted in \citet{Shikhmurzaev2020}, and it shows the influence of global flow patterns in the oscillations in the curve (which correspond to the oscillations seen in figure \ref{fig:Ucl_and_theta_c_plus_zoom}a and \ref{fig:Ucl_and_theta_c_plus_zoom}c). Figure \ref{fig:Summary_CL} summarises the process trajectory at the contact line, where dashed lines indicate the equilibrium values of the surface densities and contact angle, and arrows show the direction of advancing time. It can be clearly seen that the liquid-solid surface density at the contact line increases monotonically towards its equilibrium; while the free-surface density starts from equilibrium, falls in the vicinity of the contact line (as a consequence of the flux of mass into the liquid-solid surface). As the contact line reaches its maximum velocity, the free-surface density continues to fall for some time, then gradually returns to its equilibrium value. 

Figure \ref{fig:Summary_CL} best demonstrates the impossibility to capture the complete picture of the contact angle behaviour with a single function relating contact line velocity to contact angle, as we have two contact angles for all but one velocity. These two ``branches'' of the contact angle versus contact line velocity curve are traversed in very different time-scales, as can be seen in figure \ref{fig:Ucl_and_theta_c_plus_zoom}. Consequently, a function approximating well the relation between velocity and angle for the lower branch would still capture the greatest part of the spreading behaviour; however, it would still be lacking the physical picture of how the droplet lands on that branch.

Away from the contact line, the free-surface density remains essentially undisturbed. In the far field, the liquid-solid surface density moves, in an approximately uniform manner, towards its equilibrium, reaching equilibrium in the far field region before it does in the vicinity of the contact line.

We further consider the degenerate case of $\epsilon_g = 0$, which can be thought of as the case when the free-surface relaxation is instantaneous. This implies that the surface tension on the free surface will be constant everywhere and throughout the simulation, and from this fact it follows that there will be no slip between free surface and bulk (see equation \ref{eqn:SC1}). 

\begin{figure}
    \centering
    \includegraphics[width=\textwidth]{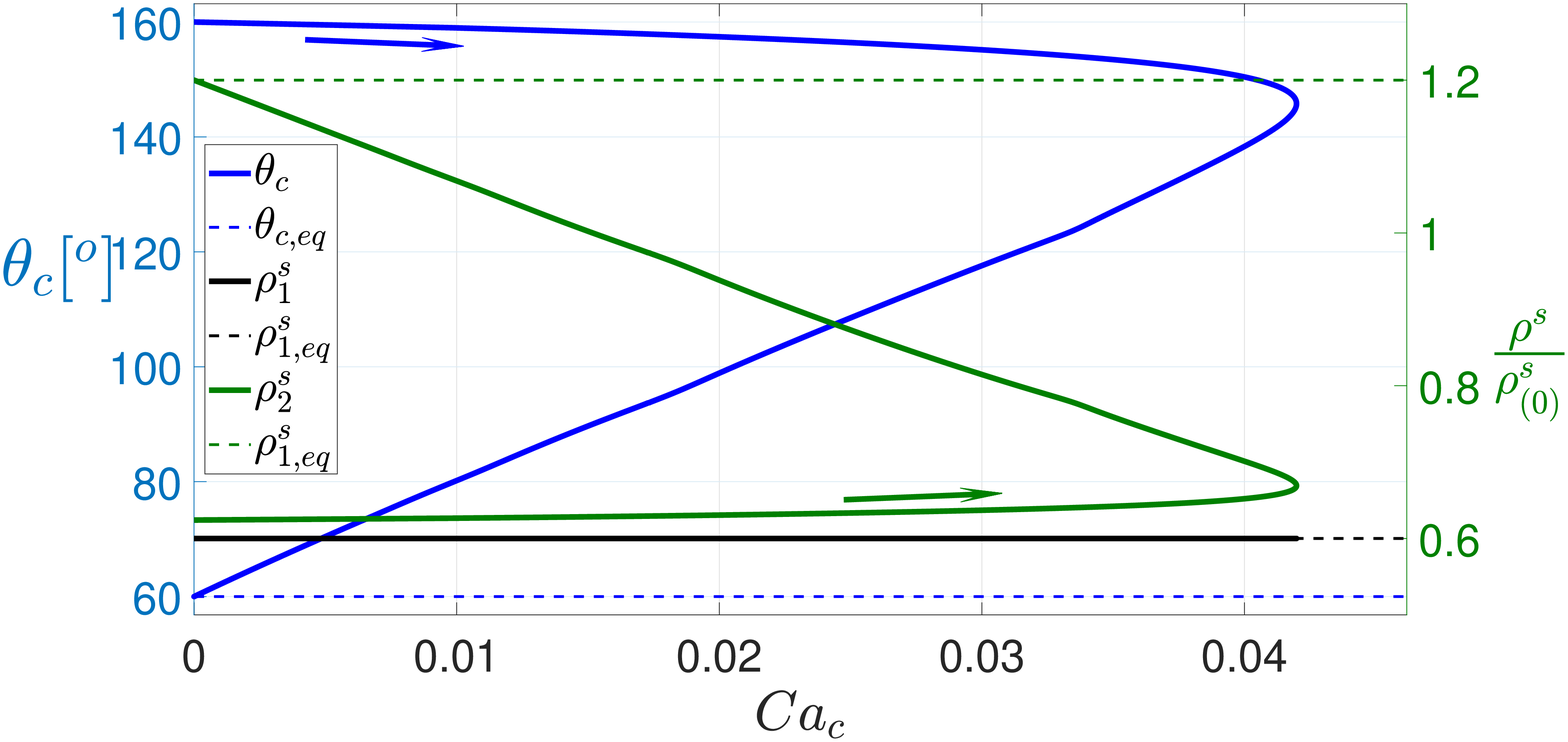}
    \caption{Process trajectories for contact angle and surface densities at the contact line in the limit case $\epsilon_g =0$.}
    \label{fig:Summary_CL_tg0}
\end{figure}

This case is of particular interest to understand the consequences for the global flow of the perturbation of the free-surface equilibrium in the vicinity of the contact line. In fact, figure \ref{fig:Summary_CL_tg0} shows that the qualitative picture is rather similar to the one shown in figure \ref{fig:Summary_CL}, with the obvious exception of the free-surface density at the contact line being kept constant. However, the maximum velocity reached by the contact line is only of about $40\%$ of that reached in the first case, which indicates that the perturbation of the free surface from its equilibrium near the contact line, contributes substantially to the mechanics of spontaneous spreading.

It should be noted that, even when the case $\epsilon_g = 0$ is essentially equivalent to the relaxation of the free surface being instantaneous, if we simply took the relaxation time of the free surface $\tau_g = 0$, coefficient $Q_g$ in equation (\ref{eqn:MEC1}) would become infinite. This issue disappears naturally with one of the improvements introduced in this work, as the dimensionless number $Q_g$ is effectively removed from the residual equations (see the discussion around equation \ref{eqn:E1_re-arranged} in the appendix for the full details). Instead, the constant $L_g =Q_g\epsilon_g =0$, which is independent of $\tau_g$, is introduced. In this improved implementation, the right-hand side of the equation controlling mass exchange between the free surface and the bulk (\ref{eqn:E1_re-arranged}) is identically null when the relaxation time is zero, as $\rho^s_1 = \rho^s$ everywhere and at all times. Nevertheless, when considering the case $\epsilon_g = 0$, it is still important to impose $L_g = 0$, to avoid undesired numerical issues. The imposition of $L_g =0$ has no negative consequences in the implementation for this case, as it does not enter the residuals elsewhere.

\section{Discussion}\label{sec:Discussion}
We implement the full set of methods, needed to model spontaneous spread of a droplet from a small initial contact area. These include, the removal of numerical artefacts known to appear when dealing with dynamic wetting with obtuse contact angles, the capture of the singular nature of pressure at the origin by the inclusion of singular elements at the contact line, and the resolution of dynamics at the scale of the interfacial layer in a continuum mechanics framework. Moreover, we use a method that produces the contact angle as an outcome of the application of first principles; which has the added advantage of allowing the contact angle to be influenced by global flow patterns. We are thus able to capture the initial regime of contact line acceleration with a decreasing contact angle,as well as the second, longer regime of the commonly modelled behaviour. 

The proper implementation of these approaches required careful design of a mesh that is suited to tackle the varied scales of the problem (ranging from the width of the interfacial layer to the radius of the droplet, and from the relaxation times of the surfaces to the duration of the spreading), while also remaining computationally tractable. 

In comparison to simpler approaches, e.g. those based on the imposition \textit{a priori} of a one-to-one map relating contact-line speed and contact angle, our model does require a substantially greater amount of work to be implemented; however, the result is well worth the effort, as cases that can not conform to the assumptions of other models are now at hand. Approaches that do not resolve the interfacial scale remain useful to model regimes where the flow results in the type of contact angle behaviour that they assume.

One difficulty with the present method is found in the implementation of the Young equation. As its derivative with respect to the contact angle is very close to zero when $\theta_c$ approaches $180^o$, thus yielding Jacobians that turn ill-posed for very large contact angles. It is possible, however, that this type of numerical issue can be made less pressing by the use of some form of preconditioning.

The main contribution of the present work is to bring together in a single implementation all the ingredients that were needed for the simulation of these spontaneous spreading flows, as well as making the necessary adaptations in each one. Each of these ingredients had been implemented independently in separate prior works \citep{SprittlesAndShikhmurzaev2011a,SprittlesAndShikhmurzaev2011b,SprittlesAndShikhmurzaev2012,SprittlesAndShikhmurzaev2013}, but they needed to be used simultaneously to deal with more general cases. Moreover, in the process of combining these methods, we were also able to improve some aspects of the implementation, as discussed in section \ref{sec:Numerical_implementation}. We highlight, in particular, our symmetric implementation of the mass balance condition for the contact line, which is introduced here and allows for the natural implementation of this boundary condition for transport along the surface, without resorting to additional assumptions on the flow of information, as was done in prior works. Further contributions include extensive notes on the derivations, which should facilitate future works to implement this model. 

In view of the experimental evidence \citep[]{RuijterEtAl1999,KarimEtAl2016,BayerAndMegaridis2006} showing the need for methods that can produce a contact angle prediction that depends on the global configuration of the flow, as well as the evidence from lattice Boltzman simulations pointing in the same direction \citep[]{DaviesEtAL2006}; it is clear that there was a missing \emph{continuum mechanics} piece of this puzzle. The present work constitutes a first fully coherent step towards providing that lacking piece in the modelling picture. Combining, adapting and improving the relevant existing strategies in a single code proved sufficient to solve the model problem at hand.

% \subsection{Future directions}
The methods used in this work are naturally relevant to most, if not all, dynamic wetting problems; thus opening countless possibilities for ramifications of the present work. More importantly, the present formulation can address problems that are known to be outside the scope of any other known method, as discussed in \citet{Shikhmurzaev2020}, while also being capable of addressing the traditional ``forced'' spreading scenarios.

Ongoing work includes the study of the dependence of the behaviour of the evolution of contact angle on physical parameters of the fluids for spreading droplets, the consideration of axisymmetric configurations, spreading over curved surfaces and spreading from a non-zero initial velocity (such as the following a droplet impact). Moreover, as the problem here considered involves a variety of scales, it is likely that its implementation can benefit from the use of different non-dimensionalisations in different parts of the domain. This and other strategies to optimise the implementation are being considered.

We highlight the relevance of the modelling here presented for drop-on-demand additive manufacturing; as this manufacturing technique involves placing individual on a substrate, where they are allow to spread and then solidify. Moreover, the framework used for this work has been shown to be capable of naturally integrating heat transfer and phase transition effects \citep{Shikhmurzaev2021,BelozerovAndShikhmurzaev2022}, making the addition of such effects a natural extension of the present work. The methods here presented are most directly applicable to the cases when drop-on-demand 3D-printing is done with Newtonian fluids, such as in the case of printing with liquid metal \citep{Simonelli2019,Gilani2021,Gilani2022}. 

\vspace{5mm}

The author is very thankful to Prof. Yulii Shikhmurzaev for the support and feedback, which were both vital for this project, and to Prof. James Sprittles for helpful discussions. The author gratefully acknowledges the support of EPSRC project EP/P031684/1.

\appendix
\section{Residual equations}\label{app:Residuals}
We introduce test functions $\phi_i$, which yield the residual equations detailed below. 

The $i$-th residual equation for momentum in the $\xi$-direction (where $\xi = x,y$) is given by
\begin{equation}
    \mathcal{M}^{\xi}_i = M^{\xi}_i + \bar{M}^{\xi}_i,
\end{equation}
where $M^{\xi}_i$ and $\bar{M}^{\xi}_i$ are the contribution of the far and near field, respectively. 

For the far field, we have
\begin{equation}
    \label{eqn:Mri_sum}
    M^{\xi}_i 
    = 
    M^{\xi,0}_i 
    +
    M^{\xi,1}_i 
    +
    M^{\xi,2}_i
    +
    M^{\xi,3}_i
    +
    M^{\xi,4}_i
\end{equation}
where
\begin{eqnarray}\label{eqn:Mr0_i}
    M^{x,0}_i 
    = 
    &&
    -
    \St
    \int
    \limits_{\Omega_{\text{f}}}
    {
    \phi_i
    \hat{g}_x
    }
    +
    \Rey
    \int
    \limits_{\Omega_{\text{f}}}
    {
    \phi_i
    \partial_t u
    }
    +
    \Rey
    \int
    \limits_{\Omega_{\text{f}}}
    {
    \left[
    \phi_i
    u
    \partial_x
    u
    +
    w
    \partial_z
    u
    \right] 
    }
    -
    \Rey 
    \int
    \limits_{\Omega_{\text{f}}}
    {
    \phi_i
    \left[
    u^c
    \partial_x
    u
    +
    w^c
    \partial_z
    u
    \right]
    }
    \nonumber
    \\
    &&
    +
    \int
    \limits_{\Omega_{\text{f}}} 
    {
    \left[
    \partial_x
    u
    \partial_x
    \phi_i
    +
    \partial_z
    u
    \partial_z
    \phi_i
    \right]
    }
    -
    \int
    \limits_{\Omega_{\text{f}}} 
    {
    p
    \partial_x
    \phi_i
    },
\end{eqnarray}
\begin{eqnarray}\label{eqn:Mz0_i}
    M^{z,0}_i 
    = 
    &&
    -
    \St
    \int
    \limits_{\Omega_{\text{f}}}
    {
    \phi_i
    \hat{g}_z
    }
    +
    \Rey 
    \int
    \limits_{\Omega_{\text{f}}}
    {
    \phi_i
    \partial_t 
    w
    }
    +
    \Rey 
    \int
    \limits_{\Omega_{\text{f}}}
    {
    \left[
    \phi_i
    u
    \partial_x
    w
    +
    w
    \partial_z
    w
    \right] 
    }
    -
    \Rey 
    \int
    \limits_{\Omega_{\text{f}}}
    {
    \phi_i
    \left[
    u^c
    \partial_x
    w
    +
    w^c
    \partial_z
    w
    \right]
    }
    \nonumber
    \\
    &&
    +
    \int
    \limits_{\Omega_{\text{f}}} 
    {
    \left[
    \partial_x
    \phi_i
    \partial_x
    w
    +
    \partial_z
    \phi_i
    \partial_z
    w
    \right]
    }
    -
    \int
    \limits_{\Omega_{\text{f}}} 
    {
    p
    \partial_z
    \phi_i
    },
\end{eqnarray}
\begin{eqnarray}\label{eqn:Mr1_i}
    M^{\xi,1}_i 
    = 
    &&
    -
    \frac{
    \sigma^1(x_{J},z_{J})
    \phi_i(x_{J},z_{J})
    }{
    \Ca
    }
    m^{1,\text{f}}_{\xi}(x_{J},z_{J})
    +
    \frac{
    \sigma^1(x_a,z_a)
    \phi_i(x_a,z_a)
    }{
    \Ca
    }
    m^{1}_{\xi}(x_a,z_a)
    \nonumber
    \\
    &&
    -
    \int
    \limits_{S_{1,\text{f}}}
    {
    \phi_i
    \left[
    n^1_x
    \partial_{\xi}
    u
    +
    n^1_z
    \partial_{\xi}
    w
    \right]
    }
    -
    \int
    \limits_{S_{1,\text{f}}}
    {
    \phi_i
    p^g
    n^1_{\xi}
    }
    +
    \frac{1}{\Ca}
    \int
    \limits_{S_{1,\text{f}}}
    {
    t^1_{\xi}
    \sigma^1
    \partial_s
    \phi_i
    },
\end{eqnarray}
\begin{eqnarray}\label{eqn:Mr2_i}
    M^{\xi,2}_i 
    = 
    &&
    -
    \int
    \limits_{S_{2,\text{f}}}
    {
    \phi_i
    \left[
    n^2_x
    \partial_{\xi}
    u
    +
    n^2_z
    \partial_{\xi}
    w
    \right]
    }
    +
    \int\limits_{S_{2,\text{f}}}
    {
    \lambda^2
    \phi_i
    n^2_{\xi}
    }
    +
    \left(
    \frac{1}{4\Ca\bar{\alpha}_s}
    +
    \bar{\beta}_s
    \right)
    \int
    \limits_{S_{2,\text{f}}}
    {
    \phi_i 
    \left[ 
    u
    t^2_x
    +
    w
    t^2_z
    \right]
    t^2_{\xi}
    }
    \nonumber
    \\
    &&
    +
    \left(
    \frac{1}{4\Ca\bar{\alpha}_s}
    -
    \bar{\beta}_s
    \right)
    \int
    \limits_{S_{2,\text{f}}}
    {
    \phi_i
    \left[
    u^s
    t^2_x
    +
    w^s
    t^2_z
    \right]
    t^2_{\xi}
    }
    -
    \frac{1}{2\Ca\bar{\alpha}_s}
    \int
    \limits_{S_{2,\text{f}}}
    { 
    \phi_i
    \left[
    u^{s}_2
    t^2_x
    +
    w^{s}_2
    t^2_z
    \right]
    t^2_{\xi}
    },
\end{eqnarray}
\begin{eqnarray}\label{eqn:Mr3_i}
    M^{\xi,3}_i 
    = 
    &&
    -
    \int
    \limits_{S_3}
    {
    \phi_i
    \left[ 
    n^3_x
    \partial_{\xi}
    u
    +
    n^3_z
    \partial_{\xi}
    w
    \right]
    }
    +
    \int\limits_{S_3}
    {
    \phi_i
    \left[ 
    \gamma^3
    t^3_{\xi}
    +
    \lambda^3
    n^3_{\xi}
    \right]
    },
\end{eqnarray}
\begin{eqnarray}\label{eqn:Mr4_i}
    M^{\xi,4}_i 
    = 
    &&
    -
    \int
    \limits_{S_4}
    {
    \phi_i
    \left[ 
    n^4_x
    \partial_{\xi}
    u
    +
    n^4_z
    \partial_{\xi}
    w
    \right]
    }
    +
    \int
    \limits_{S_4}
    {
    \phi_i
    \left[ 
    \gamma^4
    t^4_{\xi}
    + 
    \lambda^4
    n^4_{\xi}
    \right]
    },
\end{eqnarray}
where $\hat{\boldsymbol{g}} = (\hat{g}_x,\hat{g}_z)$, $\boldsymbol{u} = (u,w)$, $\boldsymbol{c} = (u^c, w^c)$, $\boldsymbol{n}_j = (n^j_x,n^j_z)$, $\boldsymbol{m}_j = (m^j_x,m^j_z)$ and $\boldsymbol{m}_1^{\text{f}} = (m^{1,\text{f}}_x,m^{1,\text{f}}_z)$ (with the $\text{f}$ super-index referring to the far-field half of the free surface), $p_g$ is the gas pressure on the free-surface, the sub-index $a$ indicates the apex of the droplet (intersection of $S_1$ and $S_3$) and the sub-index $J$ refers to the intersection of $S_1$ and $S_4$ (see figure \ref{fig:Split_domain}). Moreover, $\boldsymbol{t}_j = (t^j_x,t^j_z)$ is the unit tangent to boundary $S_j$ and $\partial_s$ stands for the derivative with respect to the arc-length variable, which increases along $S_j$ in the direction of tangent $\boldsymbol{t}_j$, and $\lambda_j\boldsymbol{n}_j$ and $\gamma_j\boldsymbol{t}_j$ are the normal and tangential stress, respectively. Moreover; $\mathbb{P}\cdot\boldsymbol{n}_i = \lambda^i\boldsymbol{n}^i+\gamma^i\boldsymbol{t}_i$, $\boldsymbol{u_s} = (u_s,w_s)$, and $\boldsymbol{v}^{s}_i = (u^s_i,w^s_i)$.

We highlight that the first integrals on the right-hand side (RHS) of equations (\ref{eqn:Mr1_i})-(\ref{eqn:Mr4_i}), though they are line integrals over the boundary, involve derivatives with respect to $x$ and $z$, and therefore the algorithm used to calculate these terms must resort to data from triangular elements in the bulk. These integrals originate from boundary terms that appear through the use of the Gauss' divergence theorem on terms that result from the divergence of the stress tensor. Bulk terms that result from this transformation are eliminated using the incompressibility condition. This results in an exact implementation of the incompressibility condition in this these bulk terms, and it has proven to eliminate a tendency to have noise in the velocities at the boundary.

For the near field, we have
\begin{equation}\label{eqn:barMx_sum}
    \bar{M}^{\xi}_i 
    =
    \bar{M}^{\xi,0}_i 
    + 
    \bar{M}^{\xi,1}_i 
    + 
    \bar{M}^{\xi,2}_i 
    +
    \bar{M}^{\xi,5}_i, 
\end{equation}
where once again we have $\xi = x,z$ and where
\begin{eqnarray}\label{eqn:barMr0_i}
    \bar{M}^{x,0}_i 
    =
    &&
    \Rey 
    \int
    \limits_{\Omega^n}
    {
    \phi_i
    \partial_t 
    u
    }
    -
    \St 
    \int
    \limits_{\Omega^n}
    {
    \phi_i
    \hat{g}_x
    }
    +
    A
    \int\limits_{\Omega^n}
    {
    \left[ 
    \partial_x
    \phi_i
    \partial_x
    \check{u}
    +
    \partial_z
    \phi_i
    \partial_z
    \check{u}
    \right] 
    }
    \nonumber
    \\
    &&
    +
    \Rey 
    \int
    \limits_{\Omega^n}
    {
    \phi_i
    \left[ 
    \bar{u}
    \partial_x
    \bar{u}
    +
    \bar{w}
    \partial_z
    \bar{u}
    -
    u^c
    \partial_x
    \bar{u}
    -
    w^c
    \partial_z
    \bar{u}
    \right] 
    }
    \nonumber
    \\
    &&
    +
    \Rey 
    A
    \int
    \limits_{\Omega^n}
    {
    \phi_i 
    \left[ 
    \check{u}
    \partial_x
    \bar{u}
    +
    \check{w}
    \partial_z
    \bar{u}
    +
    \bar{u}
    \partial_x 
    \check{u}
    +
    \bar{w}
    \partial_z
    \check{u}
    \right]
    }
    -
    \Rey
    A
    \int
    \limits_{\Omega^n}
    {
    \phi_i
    \left[
    u^c
    \partial_x
    \check{u}
    +
    w^c
    \partial_z
    \check{u}
    \right]
    }
    \nonumber
    \\
    &&
    +
    \int\limits_{\Omega^n}
    { 
    \left[
    \partial_x
    u
    \partial_x
    \phi_i
    +
    \partial_z 
    u
    \partial_z
    \phi_i
    \right]
    }
    +
    \Rey 
    \left( A\right)^2 
    \int
    \limits_{\Omega^n}
    {
    \phi_i
    \left[ 
    \check{u}
    \partial_x
    \check{u}
    +
    \check{w}
    \partial_z 
    \check{u}
    \right] 
    }
    -
    \int
    \limits_{\Omega^n}
    {
    p
    \partial_x
    \phi_i
    }, \ \ \ \ 
\end{eqnarray}
\begin{eqnarray}\label{eqn:barMz0_i}
    \bar{M}^{z,0}_i 
    =
    &&
    \Rey 
    \int
    \limits_{\Omega_{\text{n}}}
    {
    \phi_i
    \partial_t 
    w
    }
    -
    \St 
    \int
    \limits_{\Omega_{\text{n}}}
    {
    \phi_i
    \hat{g}_z
    }
    +
    A
    \int
    \limits_{\Omega_{\text{n}}}
    {
    \left[ 
    \partial_x
    \phi_i
    \partial_x
    \check{w}
    +
    \partial_z
    \phi_i
    \partial_z
    \check{w}
    \right] 
    }
    \nonumber
    \\
    &&
    +
    \Rey 
    \int
    \limits_{\Omega_{\text{n}}}
    {
    \phi_i
    \left[ 
    \bar{u}
    \partial_x
    \bar{w}
    +
    \bar{w}
    \partial_z
    \bar{w}
    -
    u^c
    \partial_x
    \bar{w}
    -
    w^c
    \partial_z
    \bar{w}
    \right] 
    }
    \nonumber
    \\
    &&
    +
    \Rey 
    A
    \int
    \limits_{\Omega_{\text{n}}}
    {
    \phi_i 
    \left[ 
    \check{u}
    \partial_x
    \bar{w}
    +
    \check{w}
    \partial_z
    \bar{w}
    +
    \bar{u}
    \partial_x 
    \check{w}
    +
    \bar{w}
    \partial_z
    \check{w}
    \right]
    }
    \nonumber
    \\
    &&
    -
    \Rey
    A
    \int
    \limits_{\Omega_{\text{n}}}
    {
    \phi_i
    \left[
    u^c
    \partial_x
    \check{w}
    +
    w^c
    \partial_z
    \check{w}
    \right]
    }
    +
    \int
    \limits_{\Omega_{\text{n}}}
    { 
    \left[
    \partial_x
    \phi_i
    \partial_x
    w
    +
    \partial_z
    \phi_i
    \partial_z 
    w
    \right]
    }
    \nonumber
    \\
    &&
    +
    \Rey 
    \left( A\right)^2 
    \int
    \limits_{\Omega_{\text{n}}}
    {
    \phi_i
    \left[ 
    \check{u}
    \partial_x
    \check{w}
    +
    \check{w}
    \partial_z 
    \check{w}
    \right] 
    }
    -
    \int
    \limits_{\Omega_{\text{n}}}
    {
    p
    \partial_z
    \phi_i
    }, \ \ \ \ 
\end{eqnarray}
\begin{eqnarray}\label{eqn:barMr1_i}
    \bar{M}^{\xi,1}_i 
    =
    &&
    \frac{
    \sigma^1(x_{J},z_{J})
    \phi_i(x_{J},z_{J})
    }{
    \Ca
    }
    m^{1,n}_{\xi}(x_{J},z_{J})
    +
    \frac{
    \sigma^1(x_c,z_c)
    \phi_i(x_c,z_c)
    }{
    \Ca
    }
    m^1_{\xi}(x_c,z_c)
    \nonumber
    \\
    &&
    -
    A
    \int
    \limits_{S_{1,\text{n}}}
    {
    \phi_i 
    \left[ 
    n^1_x
    \partial_{\xi}
    \check{u}
    +
    n^1_z
    \partial_{\xi} 
    \check{w}
    \right]
    }
    -
    \int
    \limits_{S_{1,\text{n}}}
    {
    \phi_i
    \left[
    n^1_x
    \partial_{\xi}
    \bar{u}
    +
    n^1_z
    \partial_{\xi}
    \bar{w}
    \right]
    }
    \nonumber
    \\
    &&
    -
    \int
    \limits_{S_{1,\text{n}}}
    {
    \phi_i
    p^g
    n^1_{\xi}
    }
    +
    \frac{1}{\Ca}
    \int
    \limits_{S_{1,\text{n}}}
    {
    t^1_{\xi}
    \sigma^1
    \partial_s
    \phi_i
    },
\end{eqnarray}
\begin{eqnarray}\label{eqn:barMr2_i}
    \bar{M}^{\xi,2}_i 
    =
    &&
    -
    \int
    \limits_{S_{2,\text{n}}}
    {
    \phi_i
    \left[
    A
    \left(
    n^2_r
    \partial_{\xi}
    \check{u}
    +
    n^2_z
    \partial_{\xi}
    \check{w}
    \right)
    +
    n^2_r
    \partial_{\xi}
    \bar{u}
    +
    n^2_z
    \partial_{\xi}
    \bar{w}
    \right]
    }
    \nonumber
    \\
    &&
    {
    +
    {
    \left(
    \frac{1}{4\Ca\bar{\alpha}_s}
    +
    \bar{\beta}_s
    \right)
    A
    \int
    \limits_{S_{2,\text{n}}}
    \phi_i
    \left[
    \check{u}
    t^2_r
    +
    \check{w}
    t^2_z
    \right]
    t^2_{\xi}
    }
    }
    \nonumber
    \\
    &&
    +
    \left(
    \frac{1}{4\Ca\bar{\alpha}_s}
    +
    \bar{\beta}_s
    \right)
    \int
    \limits_{S_{2,\text{n}}}
    {
    \phi_i 
    \left[
    \bar{u}
    t^2_r
    +
    \bar{w}
    t^2_z
    \right]
    t^2_{\xi}
    }
    \nonumber
    \\
    &&
    +
    \left(
    \frac{1}{4\Ca\bar{\alpha}_s}
    -
    \bar{\beta}_s
    \right)
    \int
    \limits_{S_{2,\text{n}}}
    {
    \phi_i
    \left[
    u^s
    t^2_r
    +
    w^s
    t^2_z
    \right]
    t^2_{\xi}
    }
    \nonumber
    \\
    &&
    -
    \frac{1}{2\Ca\bar{\alpha}_s}
    \int
    \limits_{S_{2,\text{n}}}
    {
    \phi_i
    \left[
    u^{s}_2
    t^2_r
    +
    w^{s}_2
    t^2_z
    \right]
    t^2_{\xi}
    }
    +
    \int
    \limits_{S_{2,\text{n}}}
    {
    \lambda^2
    \phi_i
    n^2_{\xi}
    },
\end{eqnarray}
and
\begin{eqnarray}\label{eqn:barMr4_i}
    \bar{M}^{\xi,4}_i 
    =
    &&
    A
    \int
    \limits_{S_4}
    { 
    \phi_i
    \left[
    n^4_r
    \partial_{\xi}
    \check{u}
    +
    n^4_z 
    \partial_{\xi}
    \check{w} 
    \right]
    }
    -
    \int
    \limits_{S_4}
    {
    \phi_i
    \left[
    n^4_r
    \partial_{\xi}
    \bar{u}
    +
    n^4_z
    \partial_{\xi}
    \bar{w}
    \right]
    }
    \nonumber
    \\
    &&
    -
    \int\limits_{S_4}
    {
    \phi_i
    \left[
    \lambda^4 
    n^4_{\xi}
    +
    \gamma^4
    t^4_{\xi}
    \right]
    };
\end{eqnarray}
where $\boldsymbol{\bar{u}} = (\bar{u},\bar{w})$, $\check{\boldsymbol{u}} = (\check{u},\check{w})$ and $\boldsymbol{m}^{\text{n}}_1 = (m^{1,\text{n}}_x,m^{1,\text{n}}_z)$ (where the $\text{n}$ super-index referring to the near half of the free surface). We also highlight that $\boldsymbol{n}^1$ and $\boldsymbol{n}^2$ point into the near-field domain, while $\boldsymbol{n}^4$ points out of it.

It is important to stress that the integrand in the first term on the RHS of equation (\ref{eqn:barMr0_i}) involves the full velocity of the fluid in the $x$-direction (i.e. $u = \bar{u} + A\check{u}$). In this term, it is more convenient not to separate these two components of the velocity field at this point. In this way, when the finite difference approximation for $\partial_t u$ is done, only the velocity at the time-step being calculated needs to be split into its two components, making the algorithm independent of the contribution of the eigensolution at previous time steps.

We highlight that the third integral on the RHS of equation (\ref{eqn:barMr0_i}) comes from applying Guass' divergence theorem on boundary integrals of the stress caused by the eigensolution. We note that the tangential stress caused by the eigensolution is only null along the boundaries of the wedge of angle $\theta_c$, not along the curved free surface that bounds the fluid domain.

We also note that the first term on the RHS of equation (\ref{eqn:Mr1_i}) and the last integral on the RHS of equation (\ref{eqn:Mr4_i}) will be cancelled by the analogue terms for the near field given in (\ref{eqn:barMr1_i}) and (\ref{eqn:barMr4_i}); which means that, in the computational implementation, there is no need to introduce variables $\lambda^4$, $\gamma^4$ and $\boldsymbol{m}_j(x_{J},z_{J})$, as they are cancelled in the residual equations once near-field and far-field contributions are included.

Following \citet{SprittlesAndShikhmurzaev2012}, we introduce linear test functions for the continuity equation, which we call $\psi_i$. The continuity residuals are thus given by
\begin{equation}
    C_i 
    =
    \int\limits_{\Omega_{\text{f}}}
    {
    \psi_i
    \partial_x
    u
    +
    \int\limits_{\Omega_{\text{f}}}
    \psi_i
    \partial_z
    w
    }.
\end{equation}

We note that functions $\psi_i$, which are linear in the master element coordinates, will also be used in the linear combination to expand the pressure in the bulk.

\subsection{Free surface}
We introduce surface test functions $\phi^j_i$, which coincide with the bulk test functions $\phi_i$ when sampled on boundary $S_j$, but are re-numbered along the boundary, so as to skip those which are identically zero on the interfaces.

The KBC residuals are given by $\mathcal{K}_i = K_i + \bar{K}_i$, where
\begin{equation}\label{eqn:KBC}
    K_i
    =
    \int
    \limits_{S_{1,\text{f}}}
    {
    \phi^1_{i}
    \left[ 
    n^1_x 
    \left(
    u^{s}_1
    -
    u^c
    \right)
    +
    n^1_z
    \left(
    w^{s}_1
    -
    w^c
    \right)
    \right]
    },
\end{equation} 
and the expression for $\bar{K}_i$ is identical, except for the domain of integration.

The condition for slip between the bulk and the surface phase yield the residuals $\mathcal{S}^1_i = S^1_i + \bar{S}^1_i$, where
\begin{eqnarray}
    S^{1}_i
    =
    &&
    \int
    \limits_{S_{1,\text{f}}}
    {
    \phi^1_i
    t^1_x
    \left[ 
    u^{s}_1
    -
    u
    \right]
    }
    +
    \int
    \limits_{S_{1,\text{f}}}
    {
    \phi^1_i
    t^1_z
    \left[
    w^{s}_1
    -
    w
    \right] 
    }
    +
    \frac{
    1
    +
    4
    \bar{\alpha}_g
    \bar{\beta}_g
    }{
    4
    \bar{\beta}_g
    }
    \int
    \limits_{S_{1,\text{f}}}
    {
    \sigma^{1}
    \partial_s
    \phi^1_i
    }
    \nonumber
    \\
    &&
    -
    \frac{
    1
    +
    4
    \bar{\alpha}_g
    \bar{\beta}_g
    }{
    4
    \bar{\beta}_g
    }
    \phi^1_i(x_a,z_a)
    \sigma^1(x_a,z_a)
    +
    \frac{
    1
    +
    4
    \bar{\alpha}_g
    \bar{\beta}_g
    }{
    4
    \bar{\beta}_g
    }
    \phi^1_i(x_J,z_J)
    \sigma^1(x_J,z_J)
\end{eqnarray}
and
\begin{eqnarray}
    \bar{S}^1_i
    =
    &&
    \int
    \limits_{S_{1,\text{n}}}
    {
    \phi^1_i
    \left[
    u^{s}_1
    -
    \bar{u}
    \right]
    t^1_r
    }
    +
    \int
    \limits_{S_{1,\text{n}}}
    {
    \phi^1_i
    \left[
    w^{s}_1
    -
    \bar{w}
    \right]
    t^1_z
    }
    -
    A
    \int
    \limits_{S_{1,\text{n}}}
    {
    \phi^1_i
    \check{u}
    t^1_r
    }
    -
    A
    \int
    \limits_{S_{1,\text{n}}}
    {
    \phi^1_i
    \check{w}
    t^1_z
    }
    \nonumber
    \\
    &&
    +
    \frac{
    1
    +
    4
    \bar{\alpha}_g
    \bar{\beta}_g
    }{
    4
    \bar{\beta}_g
    }\int
    \limits_{S_{1,\text{n}}}
    {
    \sigma^{1}
    \partial_s
    \phi^1_i
    }
    \nonumber
    \\
    &&
    +
    \frac{1+4\bar{\alpha}_g\bar{\beta}_g}{4\bar{\beta}_g}
    \phi^1_i(r_c,z_c)
    \sigma^1(r_c,z_c)
    -
    \frac{1+4\bar{\alpha}_g\bar{\beta}_g}{4\bar{\beta}_g}
    \phi^1_i(r_J,z_J)
    \sigma^1(r_J,z_J).
\end{eqnarray}
We note that the last term of the RHS of the two equations above will cancel each other, sa $\phi^1_i$ is continuous.

To derive the mass exchange residuals, we combine equation (\ref{eqn:MEC1}) with (\ref{eqn:DTC1}), obtaining
\begin{equation}\label{eqn:E1_re-arranged}
    \left(
    \boldsymbol{u}
    -
    \boldsymbol{v}^s_1
    \right)
    \cdot
    \boldsymbol{n}_1 
    =
    -
    L_g
    \left[
    \frac{
    \partial \rho^{s}_1
    }{
    \partial t
    }
    +
    \nabla
    \cdot
    \left(
    \rho^{s}_1
    \boldsymbol{v}^{s}_1
    \right)
    \right],
\end{equation}
where $L_g = \epsilon_g Q_g = \rho^s_{(0)}/(\rho L) $. This substitution is required as the original form of (\ref{eqn:MEC1}) yields residuals that lead to numerical issues (unnecessary operations between very small and very large numbers). These problems are naturally avoided once the equation is reformulated as shown above. Thus, using equation (\ref{eqn:DTC1_ALE}), we have $\mathcal{E}^1_i = E^1_i+\bar{E}^1_i$, where
\begin{eqnarray}\label{eqn:E1}
    E^{1}_i 
    =
    &&
    \left\{
    L_g
    \phi_i
    \rho^{s}_1
    \left[
    m^{1,\text{f}}_r
    \left(
    u_c
    -
    u^s_1
    \right)
    +
    m^{1,\text{f}}_z
    \left(
    w_c
    -
    w^{s}_1
    \right)
    \right]
    \right\}
    |_{(x_J,z_J)}
    \nonumber
    \\
    &&
    +
    L_g
    \int
    \limits_{S_{1,\text{f}}}
    {
    \phi^1_i
    \partial_t
    \rho^{s}_1
    }
    -
    L_g
    \int
    \limits_{S_{1,\text{f}}}
    {
    \rho^{s}_1
    \left(
    \partial_s
    \phi^1_i
    \right)
    \left(
    u^{s}_1
    t^1_r
    +
    w^{s}_1
    t^1_z
    \right)
    }
    \nonumber
    +
    L_g
    \int
    \limits_{S_{1,\text{f}}}
    {
    \rho^{s}_1
    \left(
    \partial_s
    \phi^1_i
    \right)
    \left(
    t^1_r
    u_c
    +
    t^1_z
    w_c
    \right)
    }
    \\
    &&
    +
    L_g
    \int
    \limits_{S_{1,\text{f}}}
    {
    \phi^1_i
    \rho^{s}_1
    \left(
    t^1_r
    \partial_s
    u_c
    +
    t^1_z
    \partial_s
    w_c
    \right)
    }
    +
    \int
    \limits_{S_{1,\text{f}}}
    {
    \phi^1_{i}
    n^1_x
    \left[ 
    u
    -
    u^{s}_1
    \right] 
    }
    + 
    \int
    \limits_{S_{1,\text{f}}}
    {
    \phi^1_{i}
    n^1_z
    \left[ 
    w
    -
    w^{s}_1
    \right]
    } \ \ 
\end{eqnarray}
and
\begin{eqnarray}
    \bar{E}^1_i 
    =
    &&
    \left\{
    L_g
    \phi_i
    \rho^{s}_1
    \left[
    m^{1,\text{f}}_r
    \left(
    u_c
    -
    u^s_1
    \right)
    +
    m^{1,\text{f}}_z
    \left(
    w_c
    -
    w^{s}_1
    \right)
    \right]
    \right\}
    |_{(x_J,z_J)}
    \nonumber
    \\
    &&
    -
    \left\{
    L_g
    \phi_i
    \rho^{s}_1
    \left[
    m^2_r
    \left(
    u_c
    -
    u^s_2
    \right)
    +
    m^2_z
    \left(
    w_c
    -
    w^{s}_2
    \right)
    \right]
    \right\}
    |_{(x_c,z_c)}
    \nonumber
    \\
    &&
    +
    L_g
    \int
    \limits_{S_{1,\text{n}}}
    {
    \phi^1_i
    \partial_t
    \rho^{s}_1
    }
    -
    L_g
    \int
    \limits_{S_{1,\text{n}}}
    {
    \rho^{s}_1
    \left(
    \partial_s
    \phi^1_i
    \right)
    \left(
    u^{s}_1
    t^1_r
    +
    w^{s}_1
    t^1_z
    \right)
    }
    \nonumber
    +
    L_g
    \int
    \limits_{S_{1,\text{n}}}
    {
    \rho^{s}_1
    \left(
    \partial_s
    \phi^1_i
    \right)
    \left(
    t^1_r
    u_c
    +
    t^1_z
    w_c
    \right)
    }
    \\
    &&
    +
    L_g
    \int
    \limits_{S_{1,\text{n}}}
    {
    \phi^1_i
    \rho^{s}_1
    \left(
    t^1_r
    \partial_s
    u_c
    +
    t^1_z
    \partial_s
    w_c
    \right)
    }
    +
    A
    \int
    \limits_{S_{1,\text{n}}}
    {
    \phi^1_{i}
    \check{u}
    n^1_x
    }
    + 
    A
    \int\limits_{S_{1,\text{n}}}
    {
    \phi^1_{i}
    \check{w}
    n^1_z
    }
    \nonumber
    \\
    &&
    \int
    \limits_{S_{1,\text{n}}}
    {
    \phi^1_{i}
    \left[
    \bar{u}
    -
    u^{s}_1
    \right]
    n^1_x
    }
    + 
    \int
    \limits_{S_{1,\text{n}}}
    {
    \phi^1_{i}
    \left[
    \bar{w}
    -
    w^{s}_1
    \right]
    n^1_z
    },
\end{eqnarray}
where we have also used the mass balance at the contact line, given by equation (\ref{eqn:MBC_ALE}), to express flows at the contact line in terms of the flux entering boundary $2$. This substitution, and others very similar to it, are, in fact, the only way in which equation (\ref{eqn:MBC_ALE}) is imposed. 

We note that the first term of each of the latter two equations (boundary contributions assessed at point $J$) will cancel each other out when the far-field and near-field contributions are added together. We also highlight that the boundary terms for flux at the apex of the droplet is omitted as it is equals to zero, which follows from equations (\ref{eqn:ICS1}) and the fact that $u^c = 0$ on boundary $3$.

The mass transport along the free surface yields 
\begin{eqnarray}\label{eqn:D1}
    \mathcal{D}^1_i
    =
    &&
    -
    \left\{
    \epsilon_g
    \phi_i
    \rho^{s}_1
    \left[
    m^2_r
    \left(
    u_c
    -
    u^s_2
    \right)
    +
    m^2_z
    \left(
    w_c
    -
    w^{s_2}_c
    \right)
    \right]
    \right\}
    |_{(x_c,z_c)}
    \nonumber
    \\
    &&
    -
    \rho^s_{1,e}
    \int
    \limits_{S_1}
    {
    \phi^1_i
    }
    +
    \int
    \limits_{S_1}
    {
    \phi^1_i
    \rho^{s}_1
    }
    +
    \epsilon_g
    \int
    \limits_{S_1}
    {
    \phi^1_i
    \partial_t
    \rho^{s}_1
    }
    -
    \epsilon_g
    \int
    \limits_{S_1}
    {
    \rho^{s}_1
    \left(
    \partial_s
    \phi^1_i
    \right)
    \left(
    u^{s}_1
    t^1_r
    +
    w^{s}_1
    t^1_z
    \right)
    }
    \nonumber
    \\
    &&
    +
    \epsilon_g
    \int
    \limits_{S_1}
    {
    \rho^{s}_1
    \left(
    \partial_s
    \phi^1_i
    \right)
    \left(
    t^1_r
    u_c
    +
    t^1_z
    w_c
    \right)
    }
    +
    \epsilon_g
    \int
    \limits_{S_1}
    {
    \phi^1_i
    \rho^{s}_1
    \left(
    t^1_r
    \partial_s
    u_c
    +
    t^1_z
    \partial_s
    w_c
    \right)
    }.
\end{eqnarray}

For simplicity, the equation above has not been divided into near-field and far-field contributions, since the expression for both are essentially identical. Moreover, had we separated the contributions, we would have had boundary terms at the separatrix, which once again would be present in both contributions, but would always cancel each other out.  Moreover, boundary terms for the apex are identically null for the same reasons as mentioned for the mass exchange residuals, and the boundary terms for the contact line in (\ref{eqn:D1}) have been written in terms of the liquid-solid surface variables, using equation (\ref{eqn:MBC_ALE}). We note that this substitution was not performed in \citet{SprittlesAndShikhmurzaev2013} nor \citet{SprittlesAndShikhmurzaev2012PoF}. However, we find that this additional substitution is necessary and sufficient to satisfy equation (\ref{eqn:MBC_ALE}) in all cases considered here.

Finally, the state equation relating surface density and surface tension along boundary $1$ is satisfied in its strong form, yielding 
\begin{equation}
    T^1_i 
    =
    \sigma^1_i
    +
    \lambda_g
    \left(
    \rho^{s}_1
    -
    1
    \right).
\end{equation}

\subsection{The liquid-solid surface}
The impermeability equation leads to the following residual
\begin{equation}\label{eqn:I_far}
    I_i 
    =
    \int
    \limits_{S_2}
    {
    \phi^2_{i}
    n^2_x
    \left(
    u^{s}_2
    -
    u_s
    \right)
    }
    + 
    \int
    \limits_{S_2}
    {
    \phi^2_{i}
    n^2_z
    \left(
    w^{s}_2
    -
    w_s
    \right)
    },
\end{equation}
where, once again there is no need to separate the integration domains, as the expression for this particular equation are identical in both domains.

The equation for slip between the bulk and the surface leads to $\mathcal{S}_i = S_i + \bar{S}_i$, where
\begin{eqnarray}
    S^2_i 
    =
    &&
    \bar{\alpha}_s
    \phi^2_i(x_L,z_L)
    \sigma^2(x_L,z_L)
    -
    \bar{\alpha}_s
    \phi^2_i(x_o,z_o)
    \sigma^2(x_o,z_o)
    \nonumber
    \\
    &&
    +
    \int
    \limits_{S_{2,\text{f}}}
    {
    \phi^2_i
    \left(
    u^{s}_2
    t^2_x
    -
    \frac{1}{2}
    u
    -
    \frac{1}{2}
    u_s
    \right)
    }
    +
    \int
    \limits_{S_{2,\text{f}}}
    {
    \phi^2_i
    \left(
    w^{s}_2 
    -
    \frac{1}{2}
    w
    -
    w_{s}
    \frac{1}{2}
    \right)
    t^2_z
    }
    \nonumber
    \\
    &&
    +
    \bar{\alpha}_s
    \int
    \limits_{S_{2,\text{f}}}
    {
    \sigma^2
    \partial_s
    \phi^2_i
    },
\end{eqnarray}
and
\begin{eqnarray}
    \bar{\mathcal{S}}^2_i 
    =
    &&
    \bar{\alpha}_s
    \phi^2_i(x_c,z_c)
    \sigma^2(x_c,z_c)
    -
    \bar{\alpha}_s
    \phi^2_i(x_L,z_L)
    \sigma^2(x_L,z_L)
    \nonumber
    \\
    &&
    +
    \int
    \limits_{S_{2,\text{n}}}
    {
    \phi^2_i
    \left(
    u^{s}_2
    -
    \frac{1}{2}
    \bar{u}
    -
    \frac{1}{2}
    u_s
    \right)
    t^2_r
    }
    +
    \int
    \limits_{S_{2,\text{n}}}
    {
    \phi^2_i
    \left(
    w^{s}_2
    -
    \frac{1}{2}
    \bar{w}
    -
    \frac{1}{2}
    w_s
    \right)
    t^2_z
    }
    \nonumber
    \\
    &&
    -
    \frac{1}{2}
    A
    \int
    \limits_{S_{2,\text{n}}}
    \check{u}
    t^2_r
    -
    \frac{1}{2}
    A
    \int
    \limits_{S_{2,\text{n}}}
    \check{w}
    t^2_z
    +
    \bar{\alpha}_s
    \int
    \limits_{S_{2,\text{n}}}
    {
    \sigma^2
    \partial_s
    \phi^2_i
    },
\end{eqnarray}
where $(x_L,z_L)$ is the intersection between boundary $2$ and the separatrix. We highlight that once again, the boundary term corresponding to the separatrix will be cancelled out whenever the test functions (in this case $\phi^2_i$) are continuous.

Following transformations that are entirely analogous to the ones performed for equation (\ref{eqn:E1}), the condition for mass exchange between surface and bulk yields $\mathcal{E}^2_i = E^2_i+\bar{E}^2_i$, where
\begin{eqnarray}
    E^{2}_i 
    =
    &&
    \left\{
    L_s
    \delta_{i,c}
    \rho^s_2
    \left[
    m^{2,\text{f}}_x
    \left(
    u_c
    -
    u^{s}_2
    \right)
    +
    m^{2,\text{f}}_z
    \left(
    w_c
    -
    w^{s}_2
    \right)
    \right]
    \right\}_{x_K,z_K}
    \nonumber
    \\
    &&
    +
    L_s
    \int
    \limits_{S_{2,\text{f}}}
    {
    \phi^2_i
    \partial_t
    \rho^s_2
    }
    -
    L_s
    \int
    \limits_{S_{2,\text{f}}}
    {
    \rho^{s_2}
    \left(
    \partial_s
    \phi^2_i
    \right)
    \left(
    u^{s_2}
    t^2_r
    +
    w^{s_2}
    t^2_z
    \right)
    }
    \nonumber
    \\
    &&
    +
    L_s
    \int
    \limits_{S_{2,\text{f}}}
    {
    \rho^{s_2}
    \left(
    \partial_s
    \phi^2_i
    \right)
    \left(
    t^2_r
    u_c
    -
    t^2_z
    w_c
    \right)
    }
    +
    L_s 
    \int
    \limits_{S_{2,\text{f}}}
    {
    \phi^2_i
    \rho^s_2
    \left(
    t^2_r
    \partial_s
    u_c
    +
    t^2_z
    \partial_s
    w_c
    \right)
    }
    \nonumber
    \\
    &&
    \int
    \limits_{S_{2,\text{f}}}
    {
    \phi^2_{i}
    n^2_x
    \left[ 
    u
    -
    u^{s}_2
    \right]
    }
    + 
    \int\limits_{S_{2,\text{f}}}
    {
    \phi^2_{i}
    n^2_z
    \left[ 
    w
    -
    w^{s}_2
    \right] 
    }
\end{eqnarray}
and
\begin{eqnarray}
    \bar{E}^2_i 
    =
    &&
    \left\{
    L_s
    \delta_{i,c}
    \rho^s_2
    \left[
    m^{2,\text{n}}_x
    \left(
    u_c
    -
    u^{s}_2
    \right)
    +
    m^{2,\text{n}}_z
    \left(
    w_c
    -
    w^{s}_2
    \right)
    \right]
    \right\}_{x_K,z_K}
    \nonumber
    \\
    &&
    -
    \left\{
    L_s
    \delta_{i,c}
    \rho^s_2
    \left[
    m^1_x
    \left(
    u_c
    -
    u^{s}_1
    \right)
    +
    m^1_z
    \left(
    w_c
    -
    w^{s}_1
    \right)
    \right]
    \right\}_{x_c,z_c}
    \nonumber
    \\
    &&
    +
    L_s
    \int
    \limits_{S_{2,\text{n}}}
    {
    \phi^2_i
    \partial_t
    \rho^s_2
    }
    -
    L_s
    \int
    \limits_{S_{2,\text{n}}}
    {
    \rho^{s_2}
    \left(
    \partial_s
    \phi^2_i
    \right)
    \left(
    u^{s_2}
    t^2_r
    +
    w^{s_2}
    t^2_z
    \right)
    }
    \nonumber
    \\
    &&
    +
    L_s
    \int
    \limits_{S_{2,\text{n}}}
    {
    \rho^{s_2}
    \left(
    \partial_s
    \phi^2_i
    \right)
    \left(
    t^2_r
    u_c
    -
    t^2_z
    w_c
    \right)
    }
    +
    L_s 
    \int
    \limits_{S_{2,\text{n}}}
    {
    \phi^2_i
    \rho^s_2
    \left(
    t^2_r
    \partial_s
    u_c
    +
    t^2_z
    \partial_s
    w_c
    \right)
    }
    \nonumber
    \\
    &&
    \int
    \limits_{S_{2,\text{n}}}
    {
    \phi^2_{i}
    \left(
    \bar{u}
    -
    u^s_2
    \right)
    n^2_r
    }
    + 
    \int
    \limits_{S_{2,\text{n}}}
    {
    \phi^2_{i}
    \left(
    \bar{w}
    -
    w^s_2
    \right)
    n^2_z
    }
    \nonumber
    \\
    &&
    +
    A
    \int
    \limits_{S_{2,\text{n}}}
    {
    \phi^2_{i}
    \check{u}
    n^2_r
    }
    + 
    A
    \int
    \limits_{S_{2,\text{n}}}
    {
    \phi^2_{i}
    \check{w}
    n^2_z
    },
\end{eqnarray}
where $L_s = \epsilon_g Q_g = \rho^s_{(0)}/(\rho L) $. Here, once again the boundary term that correspond to the separatrix will cancel each other out when we add the contributions from both sub-domains.

The condition for mass transport along the surface results in
\begin{eqnarray}
    \mathcal{D}^2_i
    =
    &&
    -
    \left\{
    \epsilon_s
    \delta_{i,c}
    \rho^s_2
    \left[
    m^1_x
    \left(
    u_c
    -
    u^{s}_1
    \right)
    +
    m^1_z
    \left(
    w_c
    -
    w^{s}_1
    \right)
    \right]
    \right\}_{x_c,z_c}
    \nonumber
    \\
    &&
    -
    \rho^s_{2,e}
    \int
    \limits_{S_2}
    {
    \phi^2_i
    }
    +
    \int
    \limits_{S_2}
    {
    \phi^2_i
    \rho^s_2
    }
    +
    \epsilon_s
    \int
    \limits_{S_2}
    {
    \phi^2_i
    \partial_t
    \rho^s_2
    }
    -
    \epsilon_s
    \int
    \limits_{S_2}
    {
    \rho^{s_2}
    \left(
    \partial_s
    \phi^2_i
    \right)
    \left(
    u^{s_2}
    t^2_r
    +
    w^{s_2}
    t^2_z
    \right)
    }
    \nonumber
    \\
    &&
    +
    \epsilon_s
    \int
    \limits_{S_2}
    {
    \rho^{s_2}
    \left(
    \partial_s
    \phi^2_i
    \right)
    \left(
    t^2_r
    u_c
    -
    t^2_z
    w_c
    \right)
    }
    +
    \epsilon_s 
    \int
    \limits_{S_2}
    {
    \phi^2_i
    \rho^s_2
    \left(
    t^2_r
    \partial_s
    u_c
    +
    t^2_z
    \partial_s
    w_c
    \right)
    }.
\end{eqnarray}
Here, in the same spirit as was done in equation (\ref{eqn:D1}), we have used equation (\ref{eqn:MBC_ALE}) to express the boundary terms as a function of the free-surface variables. 

As we did on boundary $1$, we can also satisfy the state equation in its strong form, obtaining
\begin{equation}
    T^2_i 
    =
    \sigma^2_i
    +
    \lambda_s
    \left(
    \rho^{s}_2
    -
    1
    \right).
\end{equation}

\subsection{Pressure condition}
When the contact angle is obtuse, we include $A$ as an unknown and the equation
\begin{equation}
    p(x_s,z_s) 
    -
    p(x_g,z_g) 
    =
    0,
\end{equation}
where $(x_s,z_s)$ and $(x_g,z_g)$ are the two pressure nodes, one on each interface, that are closest to the contact line, for which our mesh ensures that they be located at the same distance from the contact line.

\bibliographystyle{jfm.bst}
\bibliography{Main.bib}

\end{document}